\documentclass[onecolumn,numberedappendix,iop,apj]{emulateapj}

\usepackage{graphicx}
\usepackage{epsf}
\usepackage[title]{appendix}
\usepackage[latin2]{inputenc}

\newcommand{\vk}{\mathbf k}
\newcommand{\vp}{\mathbf p}
\newcommand{\vq}{\mathbf q}
\newcommand{\vv}{\mathbf v}
\newcommand{\vx}{\mathbf x}
\newcommand{\vu}{\mathbf u}

\newcommand{\vpsi}{\mathbf \Psi}
\newcommand{\vphi}{\mathbf \Phi}
\newcommand{\vn}{\mathbf n}

\newcommand{\mT}{\mathcal{T}}
\newcommand{\mF}{\mathcal{F}}
\newcommand{\mK}{\mathcal{K}}
\newcommand{\mH}{\mathcal{H}}
\newcommand{\mP}{\mathcal{P}}

\newcommand{\Rmnum}[1]{\uppercase\expandafter{\romannumeral #1}}

\newcommand{\psid}{\Psi_{\rho}}
\newcommand{\psiv}{\Psi_v}
\newcommand{\phid}{\Phi_{\rho}}
\newcommand{\phiv}{\Phi_v}

\newcommand{\pr}{\prime}
\newcommand{\ppr}{\prime\prime}

\newcommand{\hmpc}{\,$h^{-1}$\,Mpc}
\newcommand{\hmpcnos}{$h^{-1}$\,Mpc}
\newcommand{\ihmpc}{\,$h$\,Mpc$^{-1}$}
\newcommand{\hgpc}{\,$h^{-1}$\,Gpc}

\begin{document}
\title{Perturbation Theory of the Cosmological Log-Density Field}
\author{
Xin Wang$^{1,2,3}$, Mark Neyrinck$^{2}$, Istv\'an Szapudi$^{4}$, Alex Szalay$^{2}$,
Xuelei Chen$^{1,5}$, \\
Julien Lesgourgues$^{6,7}$, Antonio Riotto$^{6,8}$, Martin Sloth$^6$ \\ 
{\it \small 
$^1$ Key Laboratory of Optical Astronomy, National Astronomical Observatories, \\
  Chinese Academy of Sciences, Beijing 100012, China\\
$^2$ Department of Physics and Astronomy, Johns Hopkins University, Baltimore, 
    MD 21218, US \\
$^3$ Graduate School of Chinese Academy of Sciences, Beijing 100049, China\\
$^4$ Institute for Astronomy, University of Hawaii, 2680 Woodlawn Dr., HI 96822, US\\
$^5$ Center of High Energy Physics, Peking University, Beijing 100871, China\\
$^6$ CERN, PH-TH Division, CH-1211, Geneva 23, Switzerland\\
$^7$ ITP, EPFL, CH-1015 Lausanne, Switzerland.\\
$^8$ INFN, Sezione di Padova, Via Marzolo 8, I-35131 Padua, Italy }}

%
%
%

\begin{abstract}

The matter density field 
exhibits a nearly lognormal probability density distribution (PDF) after entering into the
nonlinear regime. Recently, it has been shown that the shape of the power spectrum of a logarithmically
transformed density field is very close to the linear density power spectrum, motivating an analytic study of it.
In this paper, we develop cosmological perturbation theory for the power spectrum of this field.
Our formalism is developed in the context of renormalized perturbation theory, which
helps to regulate the convergence behavior of the perturbation series, and of the Taylor-series expansion we use of the logarithmic mapping.
This approach allows us to handle the critical issue of density smoothing in a straightforward way.
We also compare our perturbative results with simulation measurements.
\end{abstract}


\section{Introduction}
The large-scale structure of the Universe is typically quantified with
the overdensity field.  This is for good reason; the overdensity field
$\delta(\vx) = \rho(\vx)/\bar{\rho}-1$ (with $\rho$ the density) seems
to be essentially a Gaussian random field at early epochs, and also
$\delta$ sources the gravitational potential in the cosmological
Poisson equation.  However, when fluctuations are large at late times
and small scales, $\delta$ acquires some inconvenient features.  It
cannot go below $-1$, but becomes arbitrarily large in non-linear
structures; this in itself forces non-Gaussianities when fluctuations
become large.

On small scales and late times, the probability density distribution
(PDF) of the density field has a roughly lognormal form.  This was
noticed regarding the galaxy density field by e.g.\ \citet{H34} and
\citet{H85}.  \citet{CJ91} explored the theoretical origins of the
roughly lognormal nature of the matter density field.  A lognormal
model for the density field shows many features of the true nonlinear
density field, such as the constraint that $\delta$ cannot go below $-1$.  As they
showed, the continuity equation can be simplified by considering a
logarithmic transformation of the density field $A(\vx,
\tau)=\ln[1+\delta(\vx,\tau)]$,
\begin{eqnarray}
 \label{eqn:logcont}
 \frac{dA(\vx,\tau)}{d\tau}=  
 \frac{d ~ \ln \bigl [1+\delta(\vx, \tau)\bigr]}{d\tau} = 
  - \nabla\cdot \vu(\vx, \tau) = - \theta(\vx, \tau).
\end{eqnarray}
Here the Lagrangian total time derivative $d/d\tau= \partial/ \partial
\tau+\vu\cdot\nabla$, $\theta=\nabla\cdot\vv$, $\tau$ is the conformal
time, and $\vu$ is the peculiar velocity.  This equation is about as
simple as one could hope for, although some complexity is hidden in
the mismatch between the Lagrangian time derivative and the Eulerian
divergence.  In the limit that $\theta$ grows according to linear
theory, given Gaussian initial conditions, this predicts an exactly
lognormal PDF for particle overdensities.  Indeed, the velocity field
remains close to Gaussian even when the density field enters the
nonlinear regime \citep[e.g.\ ][]{K94}, giving a nearly lognormal PDF
for $\delta$.

$A$ is also a natural density variable in Schr\"{o}dinger perturbation
theory \citep{SK03}, in which tree-level perturbation theory for $A$
corresponds to partial loop summation to infinite order in $\delta$.
This apparently captures the most important terms, explaining most of
the Eulerian perturbation theory loop-level variance in $\delta$.

Recently, \citet{NSS09} analyzed the power spectrum of $A$ in the
high-resolution Millennium simulation \citep{S05}.  \citet{NSS09}
found the nonlinearities in both the shape and covariance matrix of
the $A$ power spectrum to be dramatically reduced compared to the
conventional $\delta$ power spectrum.  The shape of the $A$ power
spectrum seems intriguingly close to linear theory down to $k\sim
1$\ihmpc, with $\delta$ smoothed onto a 2-\hmpcnos\ grid.  There are a
few possible reasons for this near-linearity of the $A$ power
spectrum.  It could be understandable with perturbation theory, a
possibility which we begin to explore in this paper.  It could be
linked with the suppression of density peaks in regions that have
undergone shell crossing.  Or, it could just be a coincidence of
particular aspects of a $\Lambda$CDM-like power spectrum, e.g.\ the
small-scale slope.

In this paper, we make the first perturbative prediction
for the $A$ power spectrum.  There are reasons to think
that perturbation theory for $A$ might work even better than for
$\delta$.  At small scales, the variance in $A$ is much
smaller than in $\delta$ in the nonlinear regime.  Also, it is
tempting to try to exploit the simplicity of Eq.\ (\ref{eqn:logcont})
in a Lagrangian approach.

However, working with $A$ also presents some challenges.  For the
$\delta$ field, smoothing (onto a grid, for example) has fairly
trivial effects; at large scales, it leaves phases intact and dampens
amplitudes negligibly.  But smoothing is crucial to consider for $A$,
giving different large-scale biases for different smoothing scales
\citep{NSS09}.  It is quite difficult to model final-conditions
Eulerian smoothing using Lagrangian variables; this is why we use an
Eulerian approach.  For this, we use a Taylor series expansion of the
logarithm; the slow convergence of this alternating series in itself
portends low accuracy at highly nonlinear scales where $\delta$ is
large.

Despite these challenges, the perturbative calculations we present
below achieve good agreement with simulation measurements.
In the next section, we review the basic framework of Eulerian dynamics and 
discuss the logarithmic transformation. We calculate the power spectrum of the
$A$ field in standard perturbation theory in section 3, and in renormalized
perturbation theory in section 4.
In section 5, we present our calculation and compare it with simulation, and 
we conclude in section 6.

\section{Eulerian Dynamics and the Logarithmic Transformation}

The gravitational dynamics of a pressureless fluid before shell
crossing is governed by the continuity, Euler, and Poisson 
equations.
\begin{eqnarray}
\label{eqn:dyneqn_sep}
 \frac{\partial \delta(\vx, \tau)}{\partial \tau} + \nabla \cdot 
\bigl [~ \bigl (1+\delta(\vx,\tau) \bigr) ~ \vu(\vx, \tau) ~\bigr ] 
&=& 0, \nonumber \\
\frac{\partial \vu(\vx,\tau)}{\partial \tau} + \mH(\tau) \vu(\vx, \tau) +
\vu(\vx, \tau) \cdot \nabla\vu(\vx, \tau) &=& -\nabla \Phi_N(\vx, \tau) \nonumber \\
\nabla^2 \Phi_N(\vx, \tau) = \frac{3}{2} \Omega_m(\tau) \mH^2(\tau) 
\delta(\vx, \tau). &&
\end{eqnarray}
Here, $\mH= d\ln a(\tau)/ d\ln\tau$ is the Hubble expansion rate, $a(\tau)$ is 
the scale factor,
$\Omega_m(\tau)$ is the ratio of matter density to critical density,
and $\Phi_N(\vx,\tau)$ is the Newtonian potential.
In principle, perturbation theory of the log-density field is
possible using the log-continuity equation, Eq.\ (\ref{eqn:logcont}),
together with the last two equations in Eq.\ (\ref{eqn:dyneqn_sep}).
However, we found it more straightforward to take smoothing into
account if we expand $A$ in terms of $\delta$, so this is the approach
that we take.

\subsection{Equation of Motion and Perturbation Theory}

First we briefly review results for the conventional overdensity field $\delta$.
Following \cite{CS06a,CS06b}, the equation of motion in Fourier space can be 
written in a compact form by defining the two-component variable
\begin{eqnarray}
\vpsi_a(\vk, \eta)= \bigl(\psid(\vk,\eta),~\psiv (\vk,\eta)\bigr)= 
\bigl( \delta(\vk, \eta), ~ -\theta(\vk, \eta)/\mH \bigr).
\end{eqnarray}
The index $a\in\{\rho, v\}$ stands for density or velocity
  variables, respectively.  Where numbered indices are more
  convenient, we also use $a\in\{1, 2\}$, i.e.\ $\Psi_1=\Psi_{\rho}$,
  and $\Psi_2=\Psi_v$.

The equation of motion then reads
\begin{eqnarray}
 \label{eqn:dyneqn}
 \partial_{\eta} \vpsi_a(\vk, \eta) + \Omega_{ab} \vpsi_b(\vk,\eta) = \gamma_{abc}
 (\vk,\vk_1,\vk_2) \vpsi_b(\vk_1, \eta) \vpsi_c(\vk_2, \eta),
\end{eqnarray}
with the convention that repeated Fourier arguments are integrated
over.  The time $\eta=\ln a(\tau)$ in a Einstein-de Sitter (EdS)
universe.  Here the constant matrix
\begin{eqnarray}
 \Omega_{ab} = \left[ \begin{array}{cc} 0 & -1\\ -3/2~ & 1/2 \end{array} \right ],
\end{eqnarray}
derived for an EdS universe, is still applicable in other cosmologies
with negligible corrections to the coefficients, using $\eta=\ln
D(\tau)$ (with $D$ the growth factor).

The symmetrized vertex matrix $\gamma_{abc}$ is given by
\begin{eqnarray}
\gamma_{222}(\vk,\vk_1,\vk_2) &=& \delta_D(\vk-\vk_1-\vk_2) 
\frac{|\vk_1+\vk_2|^2 (\vk_1 \cdot \vk_2)}{2 k_1^2 k_2^2} \nonumber \\
\gamma_{121}(\vk,\vk_1,\vk_2) &=& \delta_D(\vk-\vk_1-\vk_2) 
\frac{(\vk_1+\vk_2) \cdot \vk_1 }{2 k_1^2 }
\end{eqnarray}
$\gamma_{112}(\vk,\vk_1,\vk_2)=\gamma_{121}(\vk,\vk_2,\vk_1)$, and $\gamma=0$
otherwise.

Then the formal integral solution to Eq.\ (\ref{eqn:dyneqn}) can be derived as
\begin{eqnarray}
\label{eqn:solution}
\vpsi_a(\vk,\eta) = g_{ab}(\eta) \phi_b(\vk) + 
\int_0^{\eta} d\eta^{\pr} ~ g_{ab}(\eta-\eta^{\pr})
 ~\gamma_{bcd}(\vk, \vk_1, \vk_2) ~\vpsi_c(\vk_1, \eta^{\pr}) 
 \vpsi_d(\vk_2, \eta^{\pr})
\end{eqnarray}
where $\phi_a(\vk)$ denotes the initial condition 
$\phi_a(\vk)\equiv \Psi_a(\vk,\eta=0)$, and the linear propagator $g_{ab}(\eta)$
is given by
\begin{eqnarray}
 g_{ab}(\eta) = \frac{e^{\eta}}{5} \left[ \begin{array}{cc} 3 & 2\\3 & 2 
 \end{array} \right ] - \frac{e^{-3\eta/2}}{5} 
 \left[ \begin{array}{cc} -2 & 2\\3 & -3 \end{array} \right ].
\end{eqnarray}
In the following, we adopt growing-mode initial conditions 
$\phi_a(\vk) = \delta_0(\vk) u_a $.  Here $u_a=(1, 1)$ is a unit vector.

A perturbative solution to Eq.\ (\ref{eqn:solution}) is then obtained by expanding
in terms of initial fields
\begin{eqnarray}
\label{eqn:psid}
 \vpsi_a(\vk, \eta) &=& \sum_{n=1}^{\infty} \vpsi^{(n)}_a (\vk, \eta) \nonumber \\
 \vpsi^{(n)}_a(\vk, \eta) 
   &=& \int d^3\vq_{1 \cdots n} ~\delta_D(\vk-\vq_{1\cdots n})
  \mF^{(n)}_{a b_1 \cdots b_n}(\vq_1, \cdots, \vq_n;\eta) ~\phi_{b_1}(\vq_1) \cdots
  \phi_{b_n}(\vq_n) 
\end{eqnarray}
where $d^3\vq_{1 \cdots n}$ is short for $d^3\vq_1 \cdots d^3\vq_n $, and
$\vq_{1\cdots n}$ denotes $\vq_1+\cdots+\vq_n$. The kernels $\mF^{(n)}$ are fully 
symmetric functions of the wave vectors. As shown in \cite{CS06b}, they can be 
obtained in terms of $g_{ab}$ and $\gamma_{abc}$ recursively.
\begin{eqnarray}
&\mF^{(n)}_a(\vk_1,\cdots,\vk_n; \eta) \delta_D(\vk-\vk_{1\cdots n}) =
\biggl[ \sum_{m=1}^n\int_0^{\eta} d\eta^{\pr}  g_{ab}(\eta-\eta^{\pr})
~ \gamma_{bcd}(\vk, \vk_{1\cdots m}, \vk_{m+1 \cdots n}) ~\mF_c^{(m)}
(\vk_{1\cdots m}; \eta^{\pr}) \nonumber \\
&  \times ~\mF_c^{(n-m)} (\vk_{m+1\cdots n}; \eta^{\pr}) \biggr]_{\rm symmetrized}
\end{eqnarray}
For $n=1$, $\mF^{(1)}_a(\eta)= g_{ab}(\eta) u_b$.
It should be noted that in this formalism, the kernel depends on the 
time $\eta$, since it includes subleading terms in $e^{\eta}$ \citep{CS06a}. 
If one only considers the fastest-growing mode, $\mF^{(n)}_a(\eta)$ equals the well-known PT 
kernel $\exp(n \eta)\{F^{(n)}, G^{(n)}\}$.

\subsection{Logarithmic Transformation of the Field}
In general, one can define a new vector $\vphi_a$ as a nonlinear
transformation $\mT$ of $\vpsi_a$, $\vphi_a(\vk,
\eta)=\bigl(\phid(\vk, \eta), \phiv(\vk, \eta)\bigr) =\mT
\bigl[\vpsi_a(\vk, \eta) \bigr]$.  In this paper, we consider a
logarithmic transformation of the density field in configuration
space, with $\phid(\vx,\eta)=\ln [1+\psid(\vx,\eta)]$, leaving the
velocity field unchanged, i.e.\ $\phiv(\vx,\eta)=\psiv(\vx, \eta)$.

\begin{eqnarray}
\label{eqn:phid}
\phid (\vk, \eta)&=&A(\vk,\eta)=\mathcal{L} \left[A(\vx,\eta)\right] = \sum_{n=1}^{\infty}
\frac{(-1)^{n+1}}{n} \bigl[ \psid \ast \cdots \ast
\psid\bigr] (\vk, \eta) \nonumber \\
 &=& \sum_{n=1}^{\infty} \frac{(-1)^{n+1}}{n} \int d^3\vq_{1 \cdots n}
~ \delta_D(\vk-\vq_{1\cdots n})  \psid (\vq_1, \eta) \cdots 
 \psid(\vq_n, \eta),
\end{eqnarray}
where $\mathcal{L}$ represents a Fourier transform, and $\ast$ denotes a convolution of overdensity fields.
In the following, we will interchangeably use $\psid$ and $\delta$, $\phid$ and
$A$ when the meaning should be clear from context.
Substituting the perturbative series expression of $\psid(\vq, \eta)$ 
Eq.\ (\ref{eqn:psid}) into Eq.\ (\ref{eqn:phid}), we get the formal expression of
$\phid(\vk,\eta)$ in terms of the initial density field
\begin{eqnarray}
\label{eqn:phid_pert}
 \phid(\vk, \eta) &=& \sum_{n=1}^{\infty} \frac{(-1)^{n+1}}{n} 
 \int d^3\vk_{1 \cdots n} ~ \delta_D(\vk-\vk_{1\cdots n})  
  \sum_{m_1, \cdots, m_n} \int d^3\vp^1_{1\cdots m_1} \cdots 
 ~ d^3\vp^n_{1\cdots m_n} \nonumber \\ && 
\delta_D(\vk_1- \vp^1_{1\cdots m_1}) \cdots \delta_D(\vk_n- \vp^n_{1\cdots m_n})
\mF^{(m_1)}_{\rho b^1_1\cdots b^1_{m_1}}(\vp^1_1,\cdots,\vp^1_{m_1};\eta)\cdots 
\mF^{(m_n)}_{\rho  b^n_1\cdots b^n_{m_n}}(\vp^n_1,\cdots,\vp^n_{m_n};\eta)
\nonumber \\
 && \phi_{b^1_1}(\vp^1_1) \cdots \phi_{b^1_{m_1}}(\vp^1_{m_1}) \times \cdots \times
  \phi_{b^n_1}(\vp^n_1) \cdots \phi_{b^n_{m_n}}(\vp^n_{m_n}).
\end{eqnarray}

\begin{figure}[!htp]
\begin{center}
\includegraphics[width=0.9\textwidth]{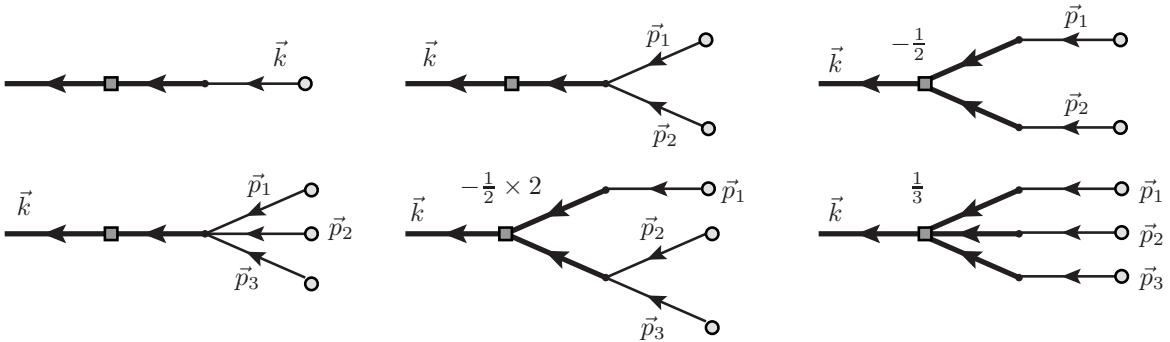}
\end{center}
\caption{\label{fig:phid} Diagrammatic representation of $\phid(\vk, \eta)$ up to 
third order. Every single branch right after the solid square represents one nonlinearly
evolved density contrast $\psid$ and is identical to, for example, diagrams at the 
right of Fig.\ 2 in \cite{BCS08}.
Explicit expressions for each diagram are shown in Eq.\ (\ref{eqn:phi_spt}) 
and  (\ref{eqn:phi_spt_win}).
}
\end{figure}

With this definition, it is possible to express $\phid(\vk)$ in a 
diagrammatic way. As depicted in Fig.(\ref{fig:phid}), each diagram contains two 
levels of interactions.  The thick solid $n$-branch tree represents the 
nonlinear convolution, i.e.\ the integration in Eq.\ (\ref{eqn:phid}), and their 
interaction vertex, shown as a solid square, carries 
a Taylor expansion coefficient $(-1)^{n+1}/n$, where $n$ is
the number of branches.
Each branch right after the grey square represents one nonlinearly evolved density contrast
$\psid(\vk_n, \eta)$ at time $\eta$, which is followed by another $m$ thin
branches, to express its gravitational nonlinearity in terms of initial fields and 
perturbative kernels $\mF^{(m)}$.
At the end of each branch, small open circles represent the initial conditions $\phi$.  
In short, in this diagrammatic representation, we just combine $n$ conventional diagrams
representing $\psid$ (e.g.\ diagrams at the right of Fig.\ 2 in \cite{BCS08}) and glue
them together at the solid square.
When expanding $\phid(\vk, \eta)$ to different orders, 
one should also include the information about the number of topologically 
equivalent diagrams, which essentially comes from the
multinomial coefficient of $(\psid)^n$. 
The final field $\phid(\vk, \eta)$ is then a summation 
of all possible diagrams.

The power spectrum $P_{\phid}(k,\eta)$  is defined as
\begin{eqnarray}
 P_{\phid}(k,\eta) \delta_D(\vk+\vk^{\pr}) = P_A(k,\eta) \delta_D(\vk+\vk^{\pr}) =  
 \langle \phid(\vk,\eta) ~ \phid(\vk^{\pr},\eta)\rangle
\end{eqnarray}
In this paper, we assume Gaussian initial conditions, with all the statistical
information encoded in the initial power spectrum
\begin{eqnarray}
\label{eqn:init_P}
 \langle \phi_a(\vk) \phi_b(\vk^{\pr}) \rangle = \delta_D(\vk+\vk^{\pr})
  P_{ab}(k),
\end{eqnarray}
where $P_{ab}(k)=u_a u_b P_0(k)$ for growing-mode initial conditions.
Diagrammatically, the ensemble average is obtained by gluing two open circles
together to form the symbol $\otimes$, which represents the initial power spectrum 
$P_{ab}(k)$.

Before proceeding, let's examine the definition of
$\phid$ carefully.  The overdensity can be idealized as a continuous field.
In practice, however, estimating a density field from points (galaxies or 
simulation particles) requires a choice of a finite resolution, effectively
a smoothing length.  
Furthermore, since the value of density field $\psid(\vx)$ strongly 
depends on 
the scale of the smoothing, as we will see, the resultant
field as well as its statistical properties also vary according to the smoothing 
procedure adopted.
In the following, we will consider the density field $\psid$ smoothed
by a spherical top-hat window function with radius $R$.

\section{Standard Perturbation Theory}
Given the perturbation expansion of $\phid(\vk, \eta)$ as well as the 
dynamics of the density field, it is straightforward to calculate the power spectrum 
of a logarithmically transformed field within the standard Eulerian perturbation theory (SPT).
Up to third order, it follows from Eq.\ (\ref{eqn:phid}) and Eq.\ (\ref{eqn:phid_pert})
that 
\begin{eqnarray}
\label{eqn:phi_spt}
\phid(k)&=& \phid^{(1)}(k) 
+ \phid^{(2)}(k) + \phid^{(3)}(k) + \cdots \nonumber \\
&=& \psid^{(1)}(k)+ \left( \psid^{(2)}(k) 
-\frac{1}{2}\bigl[\psid^{(1)} \ast \psid^{(1)} \bigr] (k) \right) +
 \biggl( \psid^{(3)}(k) -\bigl[ \psid^{(1)} \ast
 \psid^{(2)}\bigr](k)  \nonumber \\
&&  + \frac{1}{3} \left[ \psid^{(1)} \ast \psid^{(1)}\ast
\psid^{(1)}\right] (k)  \biggr) + \cdots
\end{eqnarray}
As discussed previously, a smoothing procedure is usually involved in
practice.  Applying a spherically symmetric kernel $W(kR)$ of
characteristic scale $R$, we get
\begin{eqnarray}
\label{eqn:phi_spt_win}
 \phid^{(1)}(k) &= &\psid^{(1)}(k) W(kR) \nonumber \\
 \phid^{(2)}(k) &=& \psid^{(2)}(k) W(kR) - \frac{1}{2}
  \int d^3\vp ~\psid^{(1)}(p) \psid^{(1)}(|\vk-\vp|)~ 
  W(pR) W(|\vk-\vp|R) \nonumber \\
 \phid^{(3)}(k) &=& \psid^{(3)}(k) W(kR) - \int d^3\vp 
  ~\psid^{(2)}(p) \psid^{(1)}(|\vk-\vp|) ~W(pR) W(|\vk-\vp|R)
 \nonumber \\
 && + \frac{1}{3} \int d^3\vp_{12} ~\psid^{(1)}(p_1) \psid^{(1)}(p_2) 
  \psid^{(1)}(|\vk-\vp_{12}|)~ W(p_1 R) W(p_2 R)W(|\vk-\vp_{12}|R) 
\end{eqnarray}
These expressions are shown diagrammatically in
Fig.\ \ref{fig:phid} ($\phid^{(1)}$ and $\phid^{(2)}$ in the top row, and $\phid^{(3)}$ in the bottom).  Note that the smoothing is taken over the
evolved nonlinear density field $\psid$.  The power spectrum $P_A(k)$
is then obtained by calculating the series
\begin{eqnarray}
P_A(k) = P^{(11)}_A(k) + 2 ~P^{(13)}_A(k)
 + P^{(22)}_A(k) + \cdots
\end{eqnarray}
where 
$P_A^{(ij)}(k) =  \langle \phid^{(i)}(\vk) 
\phid^{(j)}(\vk)  \rangle$.
It is clear that the tree-level power spectrum $P_A(k,\eta)$ equals exactly 
the one of the standard overdensity field 
\begin{eqnarray}
P^{(11)}_A(k,\eta)= P^{(11)}_{\delta}(k,\eta) W^2(kR) = 
D^2(\eta)  W^2(kR) P_0(k),
\end{eqnarray}
where $D(\eta)$ is the linear growth function, and $P_0(k)$ is the initial power 
spectrum. In the following, we will also denote the linear overdensity power spectrum
$P^{(11)}_{\delta}(k,\eta)$ as $P_L(k,\eta)$. 
To see the crucial role played by the smoothing window function, 
let us calculate $2P_A^{(13)}(k,\eta)$.
\begin{eqnarray}
\label{eqn:SPT_13}
  2P^{(13)}_A (k, \eta) &=&  2P^{(13)}_{\delta}(k,\eta) W^2(kR)+
  \int d^3\vp_{12} ~  \langle 
  \delta_0(\vk) \delta_0(\vp_1) \delta_0(\vp_2)  \delta_0(-\vk-\vp_{12}) 
    \rangle  \nonumber \\ && \times \biggl [ \frac{2}{3} ~ D^4(\eta)
~ W(kR)W(p_1 R) W(p_2 R) W(|\vk+\vp_{12}| R) - 
  2 \mF^{(2)}(\vp_2,-\vk-\vp_{12};\eta)  
  \nonumber \\ && \times  W(k R)W(p_1 R) W(|\vk+\vp_1| R)    \biggr],
\end{eqnarray}
where $\delta_0$ is the initial fluctuation.
This can be further simplified by performing the angular part of the 
integration. Considering only the fastest-growing mode and 
assuming spherical top-hat smoothing (see \ref{sec:1loop_spt} and \ref{sec:tophat}
for more details), Eq.\ (\ref{eqn:SPT_13}) becomes
\begin{eqnarray}
\label{eqn:SPT_13simp}
  2P^{(13)}_A (k,\eta) &=&   \left [ 2P^{(13)}_{\delta}(k,\eta)  - 
   \left(\frac{26}{21} + \frac{1}{3} \frac{d\ln \sigma^2_R}{d\ln R} \right) 
   \sigma_R^2 ~P_L (k,\eta)\right ] W^2_{\rm th}(kR)  \nonumber \\
   && - \frac{2}{3} \sigma^2_R ~ P_L(k,\eta) ~
   k R ~ W^{\prime}_{\rm th}(kR) W_{\rm th}(kR).
\end{eqnarray}
where the variance of the smoothed density fluctuation 
$\sigma^2_R = \int d^3\vq ~ P_L(q,\eta) ~ W^2_{\rm th}(qR),$ and the derivative
\begin{eqnarray}
\label{eqn:dsig}
\sigma^2_R \frac{d \ln \sigma^2_R }{d \ln R} &= &
\frac{1}{2}\int d^3\vq ~ P_L(q,\eta) ~ qR~ 
  W^{\prime}_{\rm th}(qR) W_{\rm th}(qR).
\end{eqnarray}
Here we have used the geometric properties of the spherical top-hat window function
derived by \cite{B94a,B94,B96}. 

Since $P^{(13)}_A$ is proportional to the linear power 
spectrum $P_L$, and it is well-known that  
$(P^{(13)}_{\delta}/P_L) (k\to 0) \to 0$, the power spectrum at large
scales is biased with respect to the density power spectrum
\begin{eqnarray}
\label{eqn:SPT_bias}
\frac{P_L + 2 P^{(13)}_A }{P_L}(k\to 0) = 
1 -\left(\frac{26}{21}+ \frac{1}{3} 
\frac{d\ln \sigma^2_R}{d \ln R} \right) \sigma^2_R.
\end{eqnarray}

Without the smoothing term, the bias in Eq.\ (\ref{eqn:SPT_bias}) would
equal $1-26/21~\sigma_R^2$.  This expression accords with the
first-order perturbative bias in Eq.\ (A2) (after expanding the
cumulants) of \citet{NSS09}; it is also plotted as a function of
redshift in Fig.\ 4 of that paper.  The bias becomes unphysically
negative at $z\lesssim 4$ for 2-\hmpcnos\ cells.  This smoothing term
increases the predicted bias, extending its range of
validity to significantly smaller smoothing scales.  However, we found that it still goes
pathologically negative at small smoothing scales.  This issue is
resolved in a renormalized approach, which we describe in the next
section.

\section{Renormalized Perturbation Theory}
As the density fluctuations evolve into the non-linear regime at late times, 
the validity of standard perturbation theory breaks down.  Loop 
contributions become ill-behaved and the convergence of perturbation series 
gets out of control.
Several different approaches beyond SPT have been proposed recently,
e.g.\ the renormalized perturbation theory (RPT) \citep{CS06a,CS06b,BCS08}, 
the Lagrangian resummation theory \citep{M08},
the closure theory \citep{TH08},  and the time renormalization group theory 
(TRG) \citep{P08}.
In this section, after a brief review of the RPT introduced by \cite{CS06a,CS06b},
we will show that it is possible to construct the 
perturbation series for the transformed density field based on the non-linear 
propagator of the density.

\subsection{Renormalized Perturbation Theory of the Overdensity Field }
\begin{figure}[!htp]
\begin{center}
\includegraphics[width=0.9\textwidth]{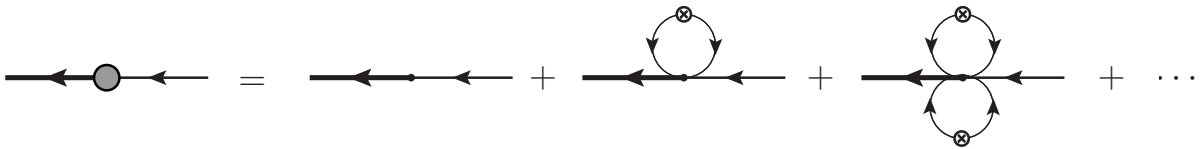}
\end{center}
\caption{\label{fig:P_psi} The nonlinear propagator $\Gamma^{(1)}_{\Psi}(k,\eta)$ 
 has an infinite number of loop contributions.
}
\end{figure}

The crucial step of renormalized perturbation theory is to define
the generalized growth factor, known as the propagator $G_{ab}(k)$
\begin{eqnarray}
\label{eqn:Gab}
  G_{ab}(k,\eta) \delta_D(\vk-\vk^{\pr}) = \Gamma^{(1)}_{\Psi~ ab}(k,\eta) \delta_D(\vk-\vk^{\pr}) \equiv \left \langle \frac{\delta \Psi_a(\vk,\eta)}
   {\delta \phi_b(\vk^{\pr})} \right \rangle,
\end{eqnarray}
which effectively describes the time evolution of individual Fourier modes when 
non-linear mode-coupling is included. 
The propagator measures the dependence of a non-linearly evolved Fourier mode 
$\Psi_a(\vk, \eta)$ on its initial state $\phi_b(\vk)$ on average.
Intuitively, one should expect $G_{ab}$ to decay to zero at small scales since 
non-linear mode-coupling has erased all the information from the initial state
at that $\vk$.
Indeed, with the help of the Feynman diagrams introduced in \cite{CS06a}, the 
dominant contribution can be summed up explicitly in the large-k limit, and giving 
the Gaussian decay $G_{ab}(k) = \exp(-k^2 \sigma_v^2 (a-1)^2/2)$. 
In this paper, the non-linear 
propagator is diagrammatically presented as a grey circle with an incoming branch.
It is a summation of infinite number of loop contributions, as illustrated in 
Fig.\ \ref{fig:P_psi}. 

Given above, \cite{CS06a} was able to rewrite the nonlinear power spectrum as 
a summation of two contributions
\begin{eqnarray}
P_{\delta}(k,\eta) = G^2(k, \eta) P_0(k) + P_{MC}(k,\eta),
\end{eqnarray}
where $G$ is the density propagator $G(k,\eta)=G_{1b}u_b$, and $P_{MC}(k,\eta)$ is 
the mode-coupling term. 
Therefore, the non-linear power spectrum at any $\vk$ is composed of two parts.
One is proportional to the initial power spectrum at the same $\vk$; the other 
comes from mode-coupling of other $\vk^{\pr}$. As $G(k)$ decays at small scales,
more and more power comes from the mode-coupling contribution.
\cite{BCS08} showed that these complicated mode-coupling contributions
can be expressed as a summation of multi-point propagators, defined as
\begin{eqnarray}
\label{eqn:gamma_psi}
\Gamma^{(n)}_{\Psi ~a b_1 \cdots b_n}  (\vk_1,\cdots,\vk_n; \eta) 
\delta_D(\vk-\vk_{1\cdots n})  = \frac{1}{n!} ~ \left \langle 
\frac{\delta^n \vpsi_a(\vk, \eta)}{\delta 
\phi_{b_1}(\vk_1) \cdots \delta \phi_{b_n}(\vk_n)  } \right \rangle,
\end{eqnarray}
which is nothing but a generalization of two-point propagator $G_{ab}$.
When concentrating on the density contrast ($a=\rho$), we will denote 
Eq.\ (\ref{eqn:gamma_psi}) as $\Gamma^{(n)}_{\delta ~ b_1\cdots b_n}$ 
(or $\Gamma^{(n)}_{\psid~ b_1\cdots b_n}$) for simplicity.
Similarly, the dominant part of multi-point propagators 
can also be summed and decay into the nonlinear regime at the same rate as 
two-point propagator.
\cite{BCS08} showed that a simple approximation which generalizes the
$k$-dependence of two-point propagators agrees with the data with acceptable 
accuracy,
\begin{eqnarray}
 \label{eqn:gammpsi_apprx}
 \Gamma^{(n)}_{\Psi} (\vk_1,\cdots,\vk_n ;\eta) = 
 \frac{\Gamma^{(1)}_{\Psi}(|\vk_{1\cdots n}|) }{\Gamma^{(1,~{\rm tree})}_{\Psi}
   (|\vk_{1\cdots n}|) } \Gamma^{(n,~{\rm tree})}_{\Psi}(\vk_1,\cdots,\vk_n;\eta)
   , \qquad (n\geq 2).
\end{eqnarray} 
With Eq.\ (\ref{eqn:gamma_psi}), the nonlinear power spectrum can then be 
expressed as
\begin{eqnarray}
\label{eqn:P_psi}
P_{\Psi~ ab}(k, \eta) = \sum_{r \ge 1} r! \int d^3\vq_{1 \cdots r} ~ 
\delta_D(\vk - \vq_{1 \cdots r}) ~
  \Gamma_{\Psi ~a }^{(r)}(\vq_1, \cdots, \vq_r; \eta) 
  \Gamma_{\Psi ~b }^{(r)}(\vq_1, \cdots, \vq_r; \eta)
 ~ P_0(q_1) \cdots P_0(q_r). \nonumber \\
\end{eqnarray}
$P_{MC}$ is the summation of all terms $r\geq 2$. 
For density fluctuations,  it reads up to one-loop order
\begin{eqnarray}
 P_{\delta}(k,\eta) = \left[  \Gamma_{\delta}^{(1)}(k;\eta) \right]^2 P_0(k) + 
 2 \int d^3\vq ~
 \left[ \Gamma_{\delta}^{(2)}(\vk-\vq, \vq; \eta) \right]^2 P_0(|\vk-\vq|) P_0(q)
\end{eqnarray}
Note that Eq.\ (\ref{eqn:P_psi}) describes nothing but an alternative way of
taking ensemble averages, or diagrammatic speaking, gluing initial states.
Instead of gluing two density fields order by order, 
one can first construct the objects by gluing initial states of individual
density fields with $n$ incoming branches, known as an ($n+1$)-point propagator, 
and the final non-linear power spectrum is then obtained by gluing two propagators 
together. 
With this observation, we can generalize this renormalization method to the 
perturbation theory of log-transformed fields.

\subsection{Renormalized Perturbation Theory of the Log-Density Field}
\begin{figure}[!htp]
\begin{center}
\includegraphics[width=0.9\textwidth]{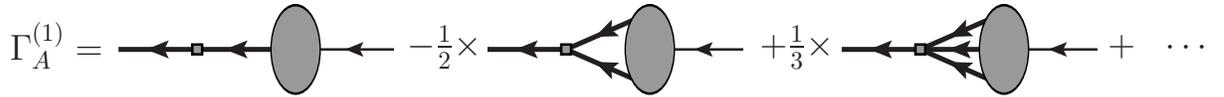}
\end{center}
\caption{\label{fig:GA1} Diagrammatic demonstration of the expansion of the 
two-point nonlinear 
propagator $\Gamma^{(1)}_A$ (Eq.\ \ref{eqn:Gamma_phi_np}). The terms on the RHS correspond to 
contributions from $n=1$ and $p=1,2$ and $3$.  }
\end{figure}

In this section, we apply the renormalized perturbation theory
Eq.\ (\ref{eqn:gamma_psi}, \ref{eqn:P_psi}) to the log-transformed field 
$\phid$.  One can define the ($n+1$)-point nonlinear propagator 
$\Gamma^{(n)}_{\Phi}$ of $\vphi_a$,
\begin{eqnarray}
\label{eqn:Gamma_phi}
\Gamma^{(n)}_{\Phi~ab_1 \cdots b_n} (\vk_1,\cdots,\vk_n; \eta) 
\delta_D(\vk-\vk_{1\cdots n})  
= \frac{1}{n!} ~\left \langle 
\frac{\delta^n \vphi_a
(\vk)}{\delta \phi_{b_1}(\vk_1) \cdots \delta \phi_{b_n}(\vk_n)  } \right \rangle.
\end{eqnarray}
For $\phid$, we will denote $\Gamma^{(n)}_{A~b_1\cdots b_n}$ or 
 $\Gamma^{(n)}_{\phid~b_1\cdots b_n}$.
Unlike its counterpart $\Gamma^{(n)}_{\Psi}$, $\Gamma^{(n)}_{\Phi}$ 
encodes not only the nonlinear information of gravitational evolution, but
also of the transformation.
Substituting the perturbation series of $\phid$ Eq.\ (\ref{eqn:phid}) into
Eq.\ (\ref{eqn:Gamma_phi}), one notices that $\Gamma^{(n)}_{\Phi}$ can be expanded 
in terms of the number of $\psid$ involved. That is, $\Gamma^{(n)}_{\Phi~ab_1 \cdots b_n} = 
\sum_p  \Gamma^{(n,~p)}_{\Phi~ab_1 \cdots b_n}$,
where 
\begin{eqnarray}
\label{eqn:Gamma_phi_np}
\Gamma^{(n,~p)}_{\Phi~ab_1 \cdots b_n} (\vk_1,\cdots,\vk_n; \eta) 
\delta_D(\vk-\vk_{1\cdots n}) &=&  \frac{(-1)^{p+1}}{n!~p}
\int d^3\vq_{1\cdots p} ~ \delta_D(\vk-\vq_{1\cdots p}) 
 \biggl \langle  \frac{\delta^n } {\delta \phi_{b_1}(\vk_1) \cdots \delta 
\phi_{b_n}(\vk_n)  }  \nonumber \\
 && \qquad \qquad \quad \left [\psid (\vq_1,\eta) \cdots \psid (\vq_p,\eta) \right ] 
 \biggr \rangle. 
\end{eqnarray}
In Fig.(\ref{fig:GA1}), we depict the expansion for the two-point propagator. The dark
ellipse here represents the complicated ensemble average in Eq.\ (\ref{eqn:Gamma_phi_np}).
Note the Taylor coefficient of the logarithmic transformation carried by each diagram.

Taking the two-point propagator as an example, now let's write down all 
contributions formally.
First of all, it is clear that the one-$\psid$ term, i.e.\ the first 
diagram in Fig.(\ref{fig:GA1}), equals $G(k, \eta)$.
For the two-$\psid$ term, i.e.\ $n=1, p=2$ in Eq.\ (\ref{eqn:Gamma_phi_np}), we can 
use Eq.\ (\ref{eqn:psid}) to further expand $\psid$ perturbatively 
\begin{eqnarray}
\label{eqn:G1_2}
\Gamma^{(1,~2)}_{A ~a}(\vk; \eta)&=& 
 -\frac{1}{2} \times 2   \int d^3 \vq_{12} ~\delta_D(\vk-\vq_{12}) 
 ~ \left \langle
 \frac{\delta \psid(\vq_1,\eta) }{ \delta \phi_a(\vk)} \psid(\vq_2, \eta) 
 \right \rangle \nonumber \\
&=& -\frac{1}{2} \times 2 \int d^3\vq_{12}
~\delta_D(\vk-\vq_{12}) \sum_{n_1, n_2}  (n_1+1)  \int d^3\vp_{1 \cdots n_1 }
 d^3\vp^{\pr}_{1 \cdots n_2} ~
  \delta_D(\vq_1-\vk-\vp_{1\cdots n_{1}} ) \nonumber \\
&&\times ~ \delta_D(\vq_2-\vp^{\pr}_{1\cdots n_2} ) 
~\mF^{(n_1+1)}_{\rho ~a c_1\cdots c_{n_1}}(\vk,\vp_1,\cdots,\vp_{n_1}; \eta) 
\mF^{(n_2)}_{\rho ~d_1\cdots d_{n_2}}(\vp^{\pr}_1,\cdots,
\vp^{\pr}_{n_2}; \eta) \nonumber \\
&& \times ~\left \langle \phi_{c_1}(\vp_1)\cdots \phi_{c_{n_1}}
(\vp_{n_1}) 
~ \phi_{d_1}(\vp^{\pr}_1) \cdots \phi_{d_{n_2}}(\vp^{\pr}_{n_2}) \right \rangle ,
\end{eqnarray}
where $n_1,~n_2$ are expansion indices of two density fluctuations $\psid$. 
The coefficient $(n_1+1)$ comes from the functional derivative
$\delta \psid/ \delta \phi_a$. Note that we have integrated out the variable
$\vp_{n_1+1}$ in the last equality together with the Dirac delta function 
$ \delta_D(\vp_{n_1+1}-\vk)$, which resulted from the functional derivative.

By Wick's theorem, the joint ensemble average in the second equality can 
be written in terms of combinations of two-point correlations.
Each of these terms can be labeled by three indices $r_1,r_2,t$, 
where $n_1=2r_1+t$, $n_2=2 r_2 +t$, i.e.\ we classify all pairs
into three categories: $r_1$ pairs within the first field 
$\phi_{c_i}(\vp_i), ~(1<i<n_1)$;
$r_2$ pairs within the second field $\phi_{d_j}(\vp^{\pr}_j), ~(1<j<n_2)$;
and $t$ pairs in between. The renormalization is then achieved by realizing 
that a pre-summation of $r_1$ and $r_2$ gives rise to one $(t+2)$-point 
and another $(t+1)$-point propagator.
Therefore, 
\begin{eqnarray}
\label{eqn:G1_2_final}
\Gamma^{(1,~2)}_{A ~a}(\vk; \eta)&=& 
 -\sum_n (n+1)! \int d^3\vp_{1 \cdots n} ~
 \Gamma^{(n+1)}_{\delta~a}(\vk,~\vp_1,\cdots,\vp_n;\eta)
 \Gamma^{(n)}_{\delta}(-\vp_1,\cdots,-\vp_n;\eta)  ~ P_0(p_1)\cdots P_0(p_n),
 \nonumber \\ 
\end{eqnarray}
where we have introduced the notation
\begin{eqnarray}
\Gamma_{\delta}^{(n)}(\vp_1,\cdots,\vp_n;\eta) \equiv 
\Gamma_{\Psi ~\rho}^{(n)}(\vp_1,\cdots,\vp_n;\eta) = 
\Gamma_{\Psi~ \rho c_1 \cdots c_n }^{(n)}(\vp_1,\cdots,\vp_n;\eta) 
 ~ u_{c_1} \cdots u_{c_n}.
\end{eqnarray}
and used the definition of $\Gamma^{(n)}_{\Psi}$ in terms of the perturbation kernel. 
The coefficient $n!$ then comes from all possible ways of matching $n$ initial
states with another group of $n$ initial states.
It is not hard to notice the similarity between Eq.\ (\ref{eqn:G1_2_final}) and
Eq.\ (\ref{eqn:P_psi}), since the derivations are nearly identical.
Yet, there are also several differences. 
For the RPT description of the matter power spectrum, both propagators share the
same order; otherwise, it would be impossible to match the pair.
For the same reason, the order of one propagator in Eq.\ (\ref{eqn:G1_2_final}) 
is always greater than the other by one, since we have to select one branch out 
before taking the average.
To conclude, we are able to construct $\Gamma_A$ from terms 
involving $\Gamma_{\delta}$.

Similarly, for the three-$\psid$ term, we have
\begin{eqnarray}
\label{eqn:G1_3}
\Gamma^{(1,~3)}_{A~ a} (\vk;\eta)  
&=&  \sum_{n_1,n_2,n_3} g_{n_1,n_2,n_3}
\int d^3\vp_{1 \cdots n_1}  d^3\vp^{\pr}_{1 \cdots n_2} 
d^3\vp^{\ppr}_{1 \cdots n_3} 
  ~\Gamma^{(n_1+n_2+1)}_{\delta~ a}(\vk; \vp_1,\cdots,\vp_{n_1}; 
  \vp_1^{\pr},\cdots,\vp^{\pr}_{n_2};\eta)  
\nonumber \\ && \times 
  ~\Gamma^{(n_2+n_3)}_{\delta}(-\vp_1^{\pr},\cdots,-\vp^{\pr}_{n_2}; 
\vp^{\ppr}_1,\cdots,\vp^{\ppr}_{n_3}; \eta ) 
~\Gamma^{(n_3+n_1)}_{\delta}( -\vp^{\ppr}_1,\cdots,-\vp^{\ppr}_{n_3}
-\vp_1,\cdots,-\vp_{n_1};\eta) 
\nonumber \\ &&\times
~ P_0(p_1) \cdots P_0(p_{n_1}) 
~ P_0(p^{\pr}_1) \cdots P_0(p^{\pr}_{n_2})
~ P_0(p^{\ppr}_1)\cdots P_0(p^{\ppr}_{n_3}). 
\end{eqnarray}
where, for each contribution
\begin{eqnarray}
 g_{n_1, n_2, n_3} &=& (n_1+n_2+1) {n_1+n_2 \choose n_1} 
  {n_2+n_3 \choose n_2}  {n_3+n_1 \choose n_3} ~ n_1! n_2! n_3! 
\end{eqnarray}
In this case, in order to describe all configurations, we need 3 indices to label the 
internal pairing and another 3 indices ($n_1, n_2, n_3$) for pairing in between.
The coefficient of each contribution $g_{n_1,n_2,n_3}$ can be understood as the number 
of equivalent  ways of gluing pairs given the triplet $\vn_c=(n_1,n_2,n_3)$.
The number of initial states for each $\psid$ is 
$\vn_v=(n_1+n_2+1, ~n_2+n_3, ~n_3+n_1)$.
After selecting one initial state from the first $\psid$ (with 
$n_1+n_2+1$ possibilities), one has to count the number of choosing 
$\vn_c$ out of $\vn_v$, which equals $ {n_1+n_2 \choose n_1} 
  {n_2+n_3 \choose n_2}  {n_3+n_1 \choose n_3} $.
The formula can be further generalized to arbitrary order $n$,
\begin{eqnarray}
\label{eqn:G1n}
\Gamma^{(1,~n)}_{A~ a} (\vk;\eta) &=& 
~ \sum_{\{t_{ij}\}}  
 g^{(1,~n)}_{\{ t_{ij} \}}
~ \biggl\{ \int~ \biggl[
\prod_{ 1\leq i<j\leq n} d^3 \vp^{ij}_{1\cdots t_{ij}} \biggl] 
~\prod_{ 1\leq i<j\leq n} \left[ P_0(p^{ij}) 
\right]^{t_{ij}} 
\nonumber \\ && \times
~\Gamma_{\delta ~a}^{(t_1 + 1)} (\vk;\vp^{11}_{1\cdots t_{11}},
\cdots, \vp^{1n}_{1\cdots t_{1n}}; \eta)
\cdots  \Gamma_{\delta}^{(t_n)} (\vp^{n1}_{1\cdots t_{n1}},\cdots,
\vp^{nn}_{1\cdots t_{nn}}; \eta ) ~  \biggr\},
\end{eqnarray}
where the index $t_{ij}$ denotes the number of connections between the $i$-th 
and the $j$-th density field $\psid$ (e.g.\ $\{t_{ij}\}=\{t_{12}, t_{13},
t_{14}, t_{23}, t_{24}, t_{34} \}$ for $n=4$), so the total number of the indices
is then $n(n-1)/2$. 
The coefficient $g^{(1,~n)}_{\{ t_{ij}\}}$ is then 
\begin{eqnarray}
\label{eqn:gtij}
g^{(1,~n)}_{\{t_{ij}\}} = (-1)^{n+1} \times ~(t_1+1)
~ \biggl[ \prod_{1\leq i\leq n} ~{t_i \choose t_{i1}\cdots t_{in} }
\biggr]  \biggl[ \prod_{1\leq i<j\leq n} (t_{ij})! \biggr].
\end{eqnarray}
Here, $t_i = \sum_{m} t_{im}$ is the total number
of connections linked between the $i$-th density field $\psid$ and others, with
$t_{ij}=t_{ji}$.
Eq.\ (\ref{eqn:gtij}) expresses the products of multinomial coefficients of 
choosing $\{t_{i1}, \cdots, t_{in}\}$ from $t_i$ for each density field 
times, all possible permutations within $t_{ij}$ for pair matching.
Each connection carries a momentum $\vp^{ij}_{l}, ~ ( 1<l<t_{ij} )$, which 
characterize the $l$-th connection between $i$-th $\psid$ and $j$-th $\psid$.
The integration is taken over momenta of all possible connections among $n$ 
different density fields. 
Because of the ensemble average Eq.\ (\ref{eqn:init_P}), we have 
$\vp^{ij}_{t_{ij}} = -\vp^{ji}_{t_{ji}}$.
And we also used the shorthand notation $ [P_0(p^{ij})]^{t_{ij}}$ for
$P_0(p^{ij}_1) \cdots  P_0(p^{ij}_{t_{ij}})$.

\begin{figure}[!htp]
\begin{center}
\includegraphics[width=0.9\textwidth]{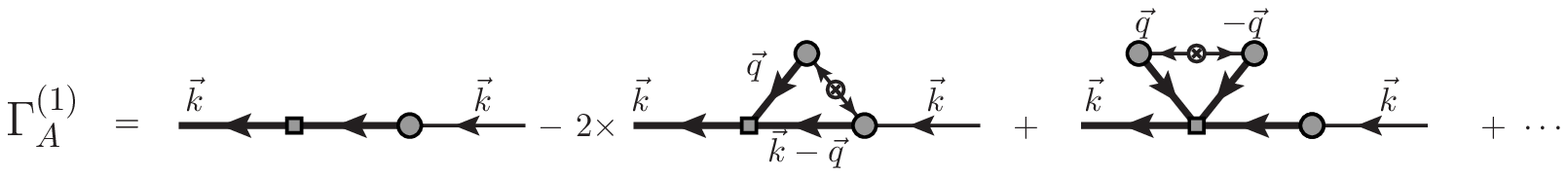}
\end{center}
\caption{\label{fig:GA1_1lp} Nonlinear propagator $\Gamma_A^{(1)}(k)$ 
of $\ln(1+\delta)$ up to one-loop order, corresponding to contributions in
Eq.\ (\ref{eqn:GA1_tree}) and (\ref{eqn:GA1_1loop}).  }
\end{figure}

\begin{figure}[!htp]
\begin{center}
\includegraphics[width=0.9\textwidth]{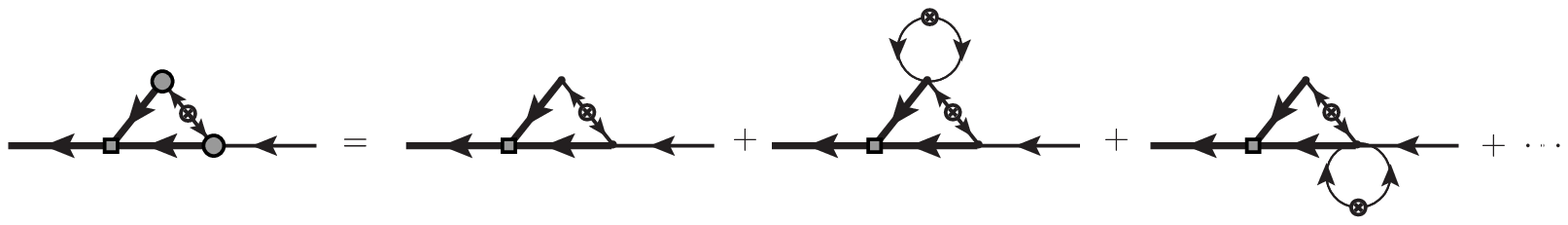}
\end{center}
\caption{\label{fig:GA1_demo} Each diagram contains an infinite number of
loop contributions.}
\end{figure}

Now we can expand $\Gamma^{(1)}_{A}$ in terms of the number of initial
power spectra entering the calculation. For the tree level, it simply reads as
\begin{eqnarray}
 \label{eqn:GA1_tree}
 \Gamma_{A ~ a}^{(1, ~{\rm tree})} (\vk;\eta) = 
 \Gamma_{\delta ~a}^{(1)}(\vk;\eta)  ~ W(kR)
\end{eqnarray}
At one-loop, from Eq.\ (\ref{eqn:G1n}), the contribution would be nonzero only when $n\le 3$,
since otherwise there would exist at least one $\Gamma_{\delta}$ with
zeroth order, which would vanish because 
$\Gamma^{(0)}_{\delta}= \langle \psid \rangle = 0$.
For $n=2$, $t_{12}=1$, the coefficient equals $-2$. For $n=3$, 
$t_{23}=1, t_{12}=t_{13}=0$, the coefficient equals $1$. So we have 
\begin{eqnarray}
\label{eqn:GA1_1loop}
 \Gamma_{A~a}^{(1, ~{\rm 1-loop})} (\vk;\eta) &= &
\int d^3\vq~P_0(q) \biggl[\left[ W(qR) ~ \Gamma_{\delta}^{(1)}(\vq;\eta) \right]^2 
~\Gamma_{\delta~a}^{(1)}(\vk;\eta) W(kR) -2~ W(qR) W(|\vk-\vq|R)~ \nonumber \\
 && \times ~ \Gamma_{\delta~a}^{(2)}(\vk, -\vq;\eta)
 \Gamma_{\delta}^{(1)}(\vq;\eta)  \biggr].
\end{eqnarray}

Given Eq.\ (\ref{eqn:G1n}), we can draw every contribution diagrammatically. 
Starting from the diagram representing $\phid^{(n)}$ in 
Fig.(\ref{fig:phid}), we change all kernels into $n$-point propagators. 
After selecting one particular branch out, we glue the rest of the initial states 
(open circles) together. 
Every resultant topologically inequivalent diagram represents one or 
several terms in Eq.\ (\ref{eqn:G1n}).
Since all ordinary kernels have already been substituted by propagators, 
the ensemble average 
(gluing) is only performed among different density fields $\psid$.
In Fig.(\ref{fig:GA1_1lp}), we show all diagrams of 
$\Gamma^{(1)}_A $ up to one-loop order.  They correspond
one-to-one to Eq.\ (\ref{eqn:GA1_tree}) and Eq.\ (\ref{eqn:GA1_1loop}).
We also present all two-loop diagrams of $\Gamma^{(1)}_A$
in Fig.(\ref{fig:GA1_2lp}).
It should be emphasized that, by substituting ordinary kernels into propagators,
each diagram in fact contains an infinite number of loop contributions at every 
substituting position, as shown in Fig.(\ref{fig:GA1_demo}).

\begin{figure}[!htp]
\begin{center}
\includegraphics[width=0.85\textwidth]{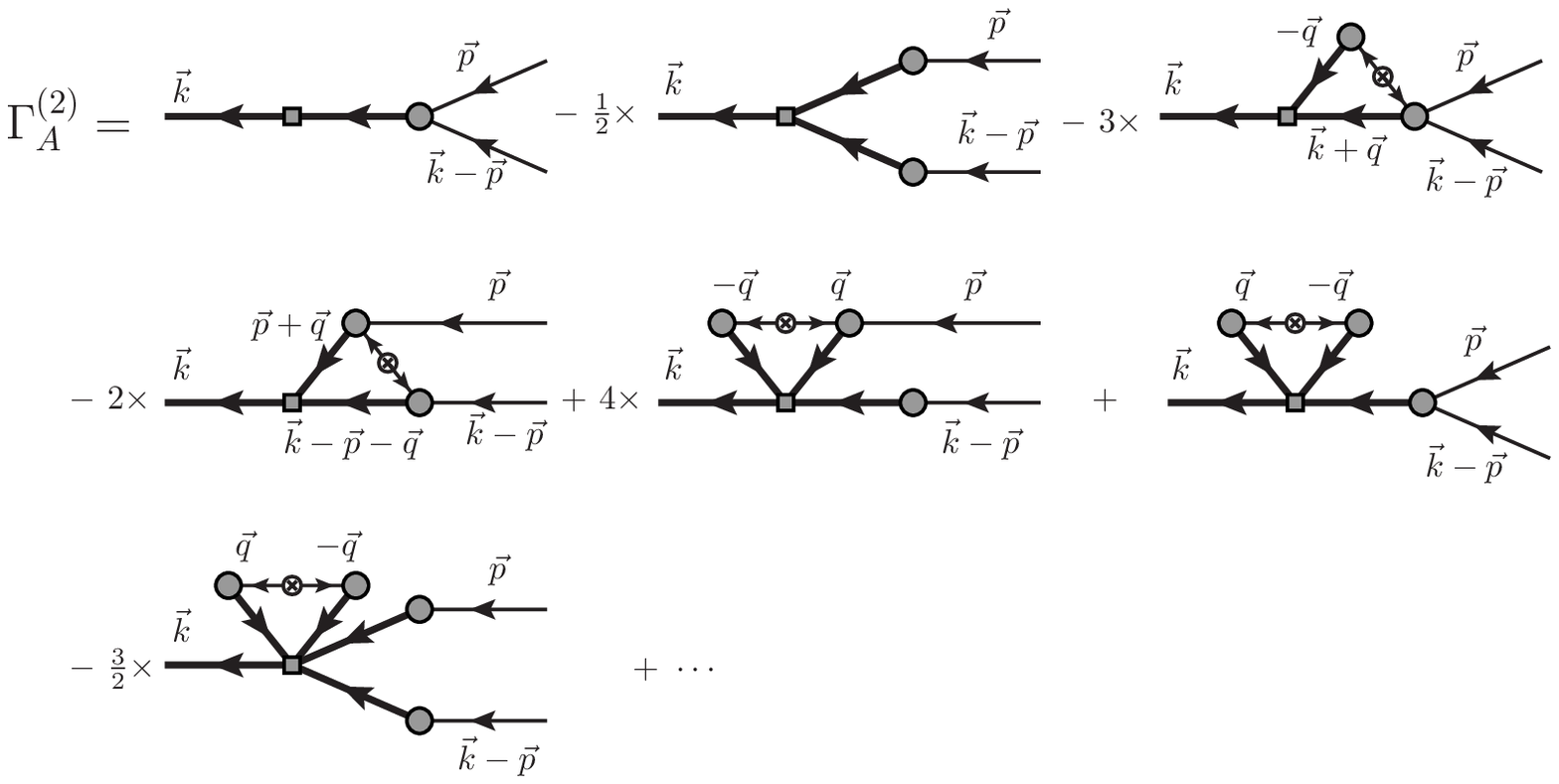}
\end{center}
\caption{\label{fig:GA2_1lp} Three-point nonlinear propagator
$\Gamma^{(2)}_A$ up to one-loop order, corresponding to contributions in 
Eq.\ (\ref{eqn:G2_tree}) and (\ref{eqn:G2_1loop}).
}
\end{figure}

For each diagram, there exists a straight path through the diagram carrying the
same momentum $\vk$ at both the start and end. Along this path, there is one
convolution vertex, as well as one $(n+1)$-point propagator $\Gamma_{\delta}^{(n)}$.
For $n=1$ (e.g.\ the third diagram in Fig.(\ref{fig:GA1_1lp})), this contribution 
recovers the same
$k$-dependence as $\Gamma_{\delta}^{(1)}(k)$, rescaled by a constant from loop 
integration.  When $n>1$, Eq.\ (\ref{eqn:gammpsi_apprx}) suggests a similar 
damping of $\vk$ given $\vp_1 \cdots \vp_{n-1}$. 
Meanwhile every propagator associates with a smoothing window function $W(kR)$.
Therefore one should expect that, in the large-$k$ limit, $\Gamma_A^{(1)}$ will decay as 
a combined effect of a Gaussian damping of $\Gamma_{\delta}^{(1)}$ and the smoothing
window function.
Furthermore, as we will see in the next section, the presence of the propagator
within the loop helps to regulate the convergence of the numerical calculation.

One can also derive the three-point propagator similarly, except that
two distinct contributions have to be taken into account because of the 
second derivative in the definition of $\Gamma^{(2)}_A$. 
\begin{eqnarray}
\label{eqn:G2_n}
\Gamma^{(2,~n)}_{A~ab} (\vk_1, \vk_2;\eta)  
&=&  \frac{(-1)^{n+1}}{2n} \times n ~ \int d^3\vq_{1\cdots n} ~
\delta_D(\vk-\vq_{1\cdots n}) ~ \biggl[ \left \langle 
\frac{\delta^2 \psid(\vq_1, \eta)} {\delta \phi_a \delta \phi_b} 
\psid(\vq_2, \eta) \cdots \psid(\vq_n, \eta) \right \rangle 
\nonumber \\
 && + ~(n-1) ~\left\langle  \frac{\delta \psid(\vq_1, \eta)} 
 {\delta \phi_a } \frac{\delta \psid(\vq_2, \eta)} 
 {\delta \phi_b } \psid(\vq_3, \eta) \cdots \psid(\vq_n, \eta) \right\rangle 
\biggr ] 
\end{eqnarray}
As shown in Eq.\ (\ref{eqn:G2_n}), the first term takes the joint average 
of the products of a second derivative with $n-1$ density fields.
Diagrammatically speaking, this category includes the first, third and sixth 
diagram in Fig.(\ref{fig:GA2_1lp}), i.e.\ both incoming branches come from 
one single density field $\psid$. 
The rest of the diagrams then correspond to the second term of Eq.\ (\ref{eqn:G2_n}),
where two branches originate from two different density fields.
Both terms can be written in the same form of Eq.\ (\ref{eqn:G1n}),
\begin{eqnarray}
\label{eqn:G2_n_detail}
\Gamma^{(2,~n)}_{A~ab} (\vk_1, \vk_2;\eta)  
&=&  \sum_{\{t_{ij}\}} ~ \int ~ \biggl[ 
\prod_{ 1\leq i<j\leq n} 
d^3 \vp^{ij}_{1\cdots t_{ij}} \biggr]
\biggl [ g^{(2,~n)}_{\{t_{ij}\}} ~
\Gamma_{\delta~ ab}^{(t_1 +2)} (\vk_1,\vk_2;\vp^{11}_{1\cdots t_{11}},
\cdots, \vp^{1n}_{1\cdots t_{1n}})
~ \times \cdots  \nonumber \\
&& \times ~ \Gamma_{\delta }^{(t_n)} (\vp^{n1}_{1\cdots t_{n1}},\cdots,
\vp^{nn}_{1\cdots t_{nn}} ) ~
  + ~\tilde{g}^{(2,~n)}_{\{t_{ij}\}} ~
\Gamma_{\delta~ a}^{( t_1+1)} (\vk_1;\vp^{11}_{1\cdots 
t_{11}}, \cdots, \vp^{1n}_{1\cdots t_{1n}}) \nonumber \\
&& \times ~ \Gamma_{\delta~ b}^{( t_2+1)} (\vk_2;\vp^{21}_{1\cdots 
t_{21}}, \cdots, \vp^{2n}_{1\cdots t_{2n}})
\cdots  \Gamma_{\delta}^{(t_n)} (\vp^{n1}_{1\cdots t_{n1}},\cdots,
\vp^{nn}_{1\cdots t_{nn}} ) \biggr ] 
\prod_{1\leq i<j\leq n} \left[ P_0(p^{ij})  \right]^{t_{ij}} 
\nonumber \\
\end{eqnarray}
The difference between the two terms can be clearly seen from their orders of their
propagators.  With the same labeling system, the first gives $\Gamma^{(t_1+2)}_{\delta} 
\Gamma^{(t_2)}_{\delta} \cdots \Gamma^{(t_n)}_{\delta}$, while the second
gives $\Gamma^{(t_1+1)}_{\delta} \Gamma^{(t_2+1)}_{\delta} \cdots 
\Gamma^{(t_n)}_{\delta}$.
Meanwhile, the two $g$ coefficients equal
\begin{eqnarray}
 g^{(2,~n)}_{\{t_{ij}\}}& =&  \frac{(-1)^{n+1}}{2} \times (t_1 +2)(t_1 +1) ~
\biggl[ \prod_{1\leq i\leq n} ~ {t_i \choose t_{i1}\cdots t_{in} }\biggr] 
 ~\biggl[\prod_{1\leq i<j\leq n} (t_{ij})! \biggr], \nonumber \\
 \tilde{g}^{(2,~n)}_{\{t_{ij}\}}& =& \frac{(-1)^{n+1}}{2} \times (n-1) (t_1+1)
 (t_2 +1) ~ \biggl[ \prod_{1\leq i\leq n} ~ {t_i \choose t_{i1}\cdots t_{in} }\biggr] 
 ~\biggl[\prod_{1\leq i<j\leq n} (t_{ij})! \biggr].
\end{eqnarray}
The first contribution $g^{(2,~n)}_{\{t_{ij}\}}$ gives $n (t_1+2)(t_1+1)$, 
while the second term $ \tilde{g}^{(2,~n)}_{\{t_{ij}\}}$ gives $ n(n-1)(t_1+1)(t_2+1)$.

At tree level, there are two diagrams: one is
$\Gamma^{(2)}_{\delta}$, from the single-$\psid$ contribution;
and the other is the two-$\psid$ term with $t_{12}=0$ for the second term of 
Eq.\ (\ref{eqn:G2_n_detail}),
\begin{eqnarray}
\label{eqn:G2_tree}
\Gamma^{(2,~{\rm tree})}_{A~ab} (\vp,\vk-\vp; \eta) = 
 \Gamma_{\delta~ab}^{(2)}(\vp,\vk-\vp;\eta)  ~ W(kR) -\frac{1}{2} 
\Gamma_{\delta~a}^{(1)}(\vp; \eta) 
\Gamma_{\delta~b}^{(1)}(\vk-\vp; \eta) ~ W(pR)W(|\vk-\vp|R). 
\end{eqnarray}
At one-loop level, five diagrams are nonzero.
\begin{eqnarray}
\label{eqn:G2_1loop}
\Gamma^{(2,~{\rm 1-loop})}_{A~ab} (\vp,\vk-\vp; \eta) &= &
\int d^3\vp ~P_0(q) ~ \left[ 
 \sum_{i=1}^5 ~ \mK^{(\rm 2,~1-loop)}_i
\right].
\end{eqnarray}
where 
\begin{eqnarray}
\label{eqn:G2_1loop_ker}
\mK^{(\rm 2,~1-loop)}_1 &=&-3 ~ W(qR) W(|\vk+\vq|R) 
~\Gamma^{(1)}_{\delta}(-\vq;\eta) 
\Gamma^{(3)}_{\delta~ab}(\vk-\vp, \vp, \vq;\eta) \nonumber \\
\mK^{(\rm 2,~1-loop)}_2 &=&  -2~  W(|\vp+\vq|R) W(|\vk-\vp-\vq|R) 
~  \Gamma^{(2)}_{\delta~a}(\vp,\vq;\eta) 
 \Gamma^{(2)}_{\delta~b}(\vk-\vp,-\vq;\eta)  \nonumber \\
\mK^{(\rm 2,~1-loop)}_3 &=& ~ 4 W(qR)W(|\vp+\vq|R) W(|\vk-\vp|R)  
~  \Gamma^{(1)}_{\delta}(\vq;\eta) \Gamma^{(1)}_{\delta~a}(\vk-\vp;\eta) 
 \Gamma^{(2)}_{\delta~b}(\vp, \vq;\eta) \nonumber \\
\mK^{(\rm 2,~1-loop)}_4 &=& ~ \left[ W(qR) \Gamma^{(1)}_{\delta}(\vq;\eta) \right]^2 
~ W(kR) \Gamma^{(2)}_{\delta~ab}(\vp, \vk-\vp;\eta) \nonumber \\
\mK^{(\rm 2,~1-loop)}_5 &=&  ~- \frac{3}{2} W^2(qR) W(|\vk-\vq|R) W(kR) 
  ~ \left[\Gamma^{(1)}_{\delta}(\vq;\eta)\right]^2 
   \Gamma^{(1)}_{\delta~a}(\vk-\vq;\eta) \Gamma^{(1)}_{\delta~b}(\vp;\eta).
\end{eqnarray}

In Eq.\ (\ref{eqn:G2_1loop_ker}), $\mK_1$ and $\mK_2$ are two-$\psid$ 
contributions from the first and second terms of Eq.\ (\ref{eqn:G2_n}) respectively,
$\mK_3$ and $\mK_4$ are three-$\psid$ contributions, and $\mK_5$ is the
four-$\psid$ contribution.

At last, the non-linear power spectrum of the log-transformed field 
$\phid$ can be expressed as 
\begin{eqnarray}
\label{eqn:P_phi}
P_A(k;\eta) = \sum_{r \ge 1} r! \int d^3\vq_{1 \cdots r} ~ 
\delta_D(\vk - \vq_{1 \cdots r}) 
 \left [ \Gamma_A^{(r)}(\vq_1, \cdots, \vq_r; \eta)  \right ]^2 
 P_0(q_1) \cdots P_0(q_r).
\end{eqnarray}

\section{Numerical Results}
In this section, we show our numerical results. 
To simplify the calculation, we adopt the approximation 
Eq.\ (\ref{eqn:gammpsi_apprx}), which has been shown to be accurate 
enough in \cite{BCS08}.
For the tree-level propagator, we only include the fastest-growing mode, i.e.
the standard perturbation kernel $F^{(n)}$.
Since the non-linear quantity $\Gamma^{(n)}_{\delta}$ is involved
throughout our formulae, all the calculation is done numerically.  
All results assume the fiducial concordance $\Lambda$CDM cosmology of
simulation 0 of the Coyote Universe suite \citep{cu1,cu2,cu3}, with
($\Omega_m h^2$, $\Omega_b h^2$, $n_s$, $w$, $\sigma_8$, $h$) =
(0.1296, 0.0224, 0.97, -1, 0.8, 0.72).  We use these parameters because we
compare our perturbative results to measurements from this simulation.
The linear power spectrum is calculated with the public Boltzmann code {\scshape camb} \citep{camb}, 
and the numerical integration is performed with the {\scshape cernlib} multi-dimensional integration routine.

\begin{figure*}[!htp]
\begin{center}
\includegraphics[width=0.45\textwidth]{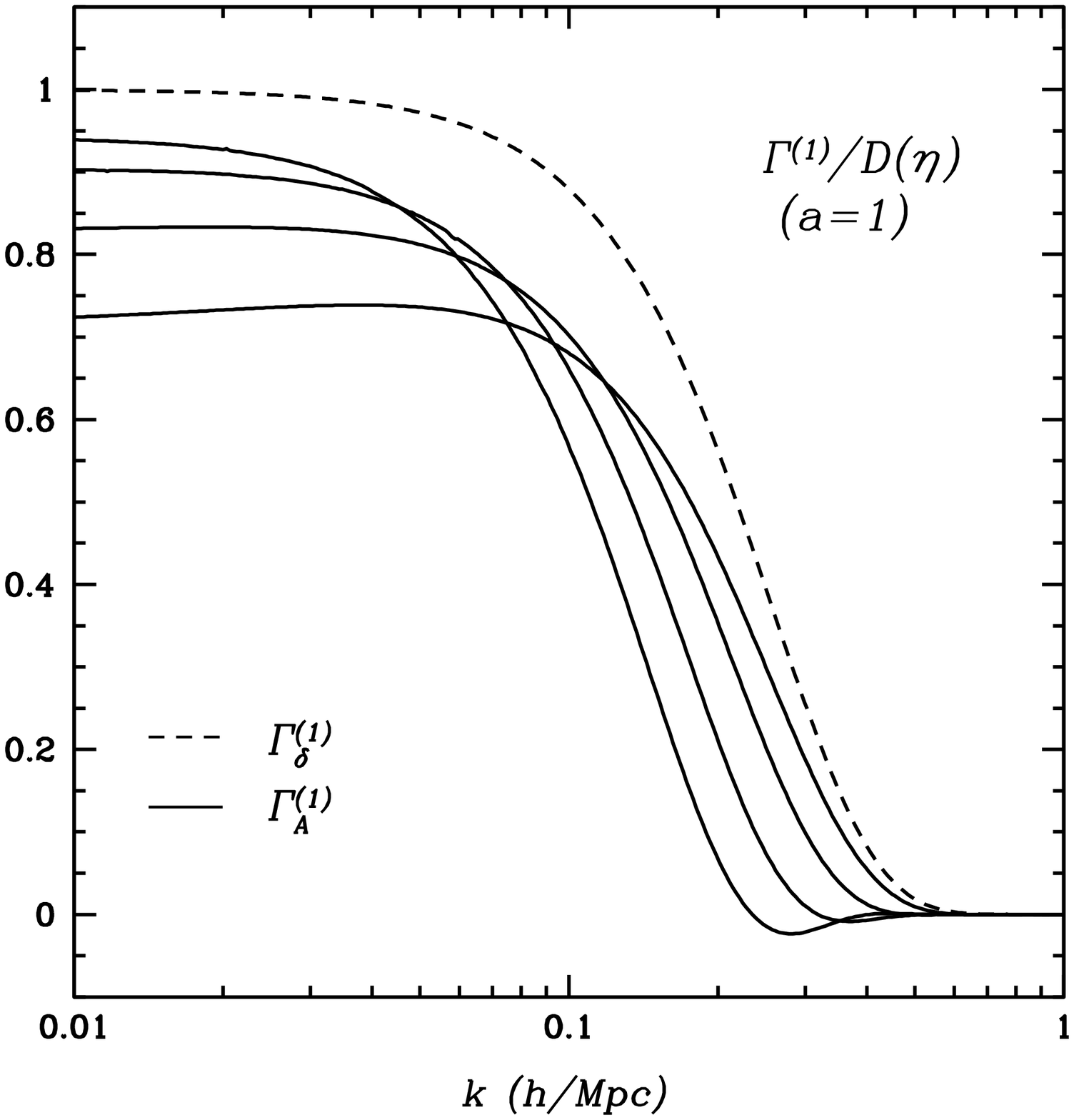}
\includegraphics[width=0.45\textwidth]{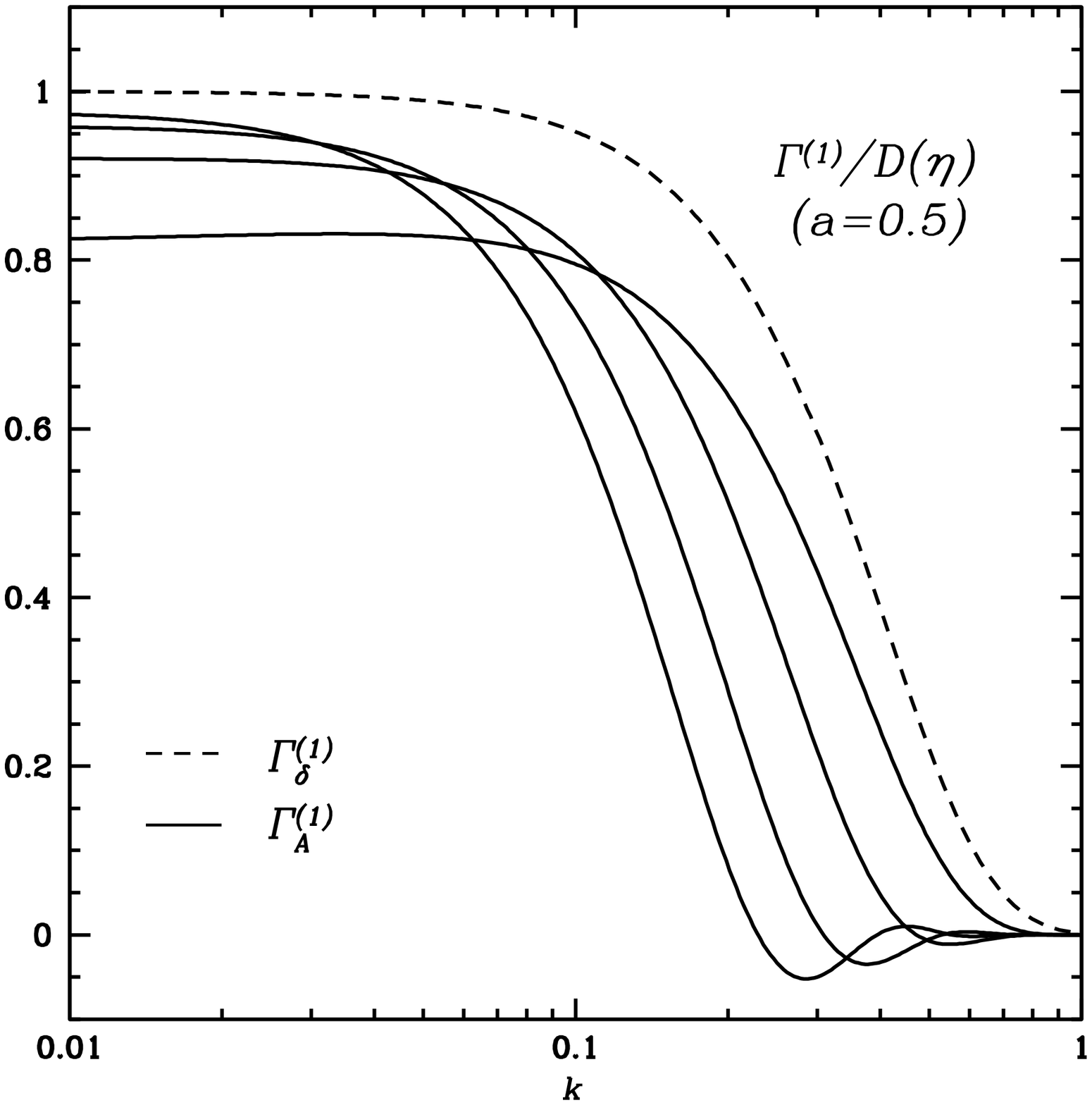}
\caption{\label{fig:gamma_phi}  Non-linear propagator over linear growth
rate at $a=1$ and $a=0.5$. 
The dashed line shows the two-point propagator of the overdensity
field $\Gamma^{(1)}_{\delta}$. Solid lines show the propagator for the
log-transformed field $\Gamma^{(1)}_A$, with top-hat 
smoothing radius $R=20$\hmpc, $15$\hmpc,
$10$\hmpc\ and $5$\hmpc, from top to bottom.
}
\end{center}
\end{figure*}

In Fig.(\ref{fig:gamma_phi}), we show the two-point propagator 
$\Gamma^{(1)}_A(k)$ up to two-loop order.
As expected from the analysis of SPT, the propagator of the $A$ field
approaches a constant less than one at large scales. 
The value of this large-scale bias, which also encodes the statistical information 
of the density fluctuation (Eq.\ref{eqn:SPT_bias}), 
strongly depends on the smoothing process adopted before the transformation.
For the smoothing radius $R \to \infty$, effectively no loop integration 
contributes to $\Gamma_A^{(1)}(k)$, therefore the bias approaches to $1$ from the
tree-level result, Eq.\ (\ref{eqn:GA1_tree}). 
When the smoothing radius goes smaller, since $\Gamma_{\delta}^{(n)}(k)$ that appears in the 
loop integration also decays at large $k$, the effect of the smoothing 
becomes less and less important until the bias freezes at some value, where higher-loop 
contributions are significant.
In this sense, the presence of $\Gamma_{\delta}^{(n)}(k)$ helps to regulate the 
convergence of the perturbative series.

At small scales, $\Gamma_A^{(1)}(k)$ is dampened both by nonlinearities and by smoothing, 
while $\Gamma_{\delta}^{(1)}(k)$ has no additional smoothing imposed.  This is why $\Gamma_A^{(1)}(k)$ decays at larger scales than $\Gamma_{\delta}^{(1)}(k)$ for large smoothing scales.  The top solid line shows
$\Gamma^{(1)}_A(k)$ with a smoothing scale $R=20$\hmpc,
which reduces to 0.5 around $k\sim 0.1$\ihmpc.
As the smoothing radius becomes smaller, the curve approaches
$\Gamma_{\delta}^{(1)}(k)$ at large $k$.
For $R=5$\hmpc\ (the bottom solid 
line), the damping of $\Gamma_A^{(1)}(k)$ is similar to that of $\Gamma_\delta^{(1)}(k)$.

If the large-scale bias is divided out, boosting the $A$ propagators
to line up for small $k$, the $A$ propagator may slightly exceed the
$\delta$ propagator on small scales, if the smoothing scale is
sufficiently small.  We have found this to be the case in preliminary
simulation measurements.  Disappointingly, this implies that the
logarithmic transform by itself does not help appreciably to
reconstruct mode-by-mode initial phases and amplitudes.  But the
similarity in damping is not surprising, given that e.g.\ bulk
displacements from the initial conditions affect both fields.

\begin{figure*}[!htp]
\begin{center}
\includegraphics[width=0.45\textwidth]{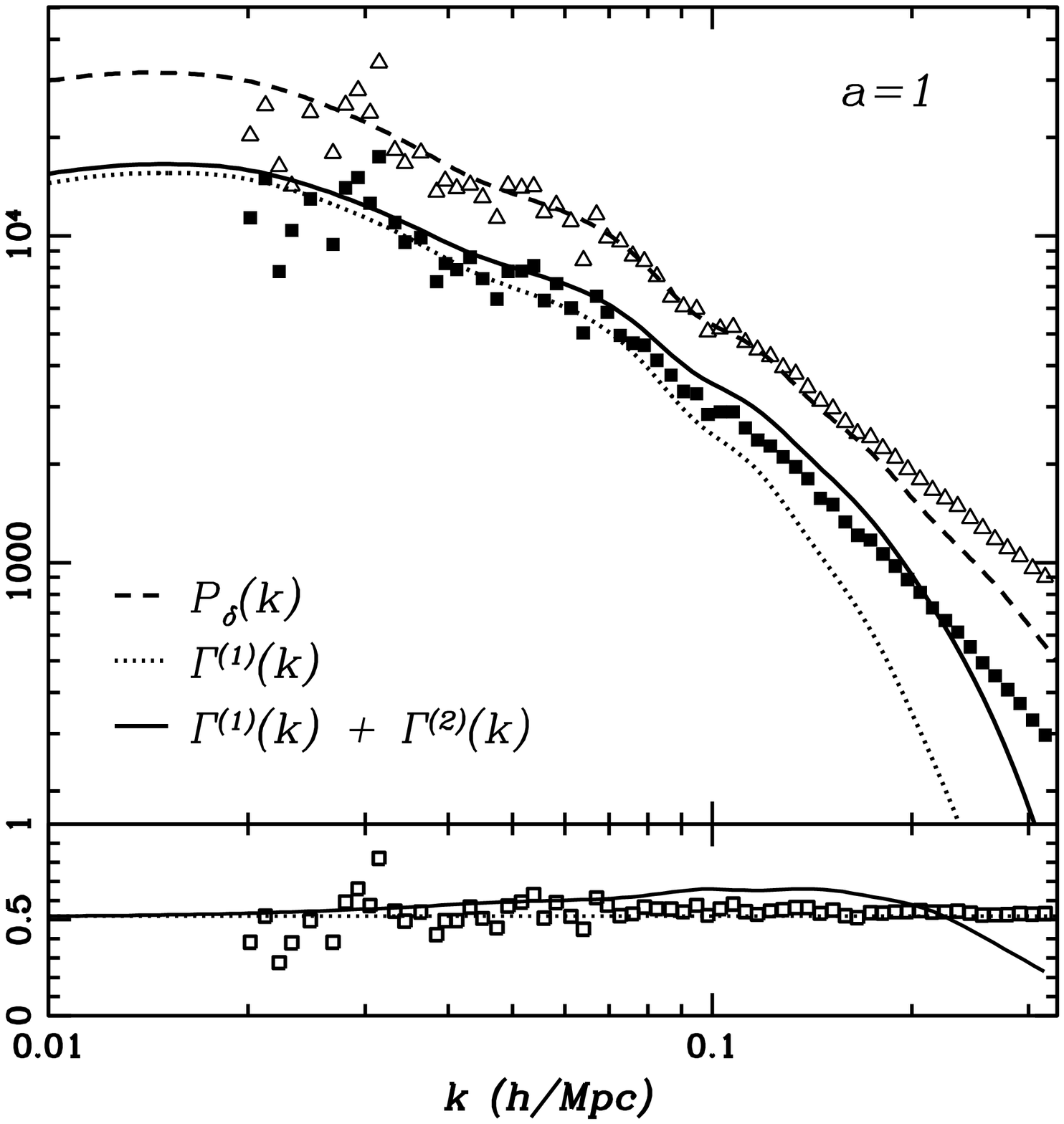}
\includegraphics[width=0.45\textwidth]{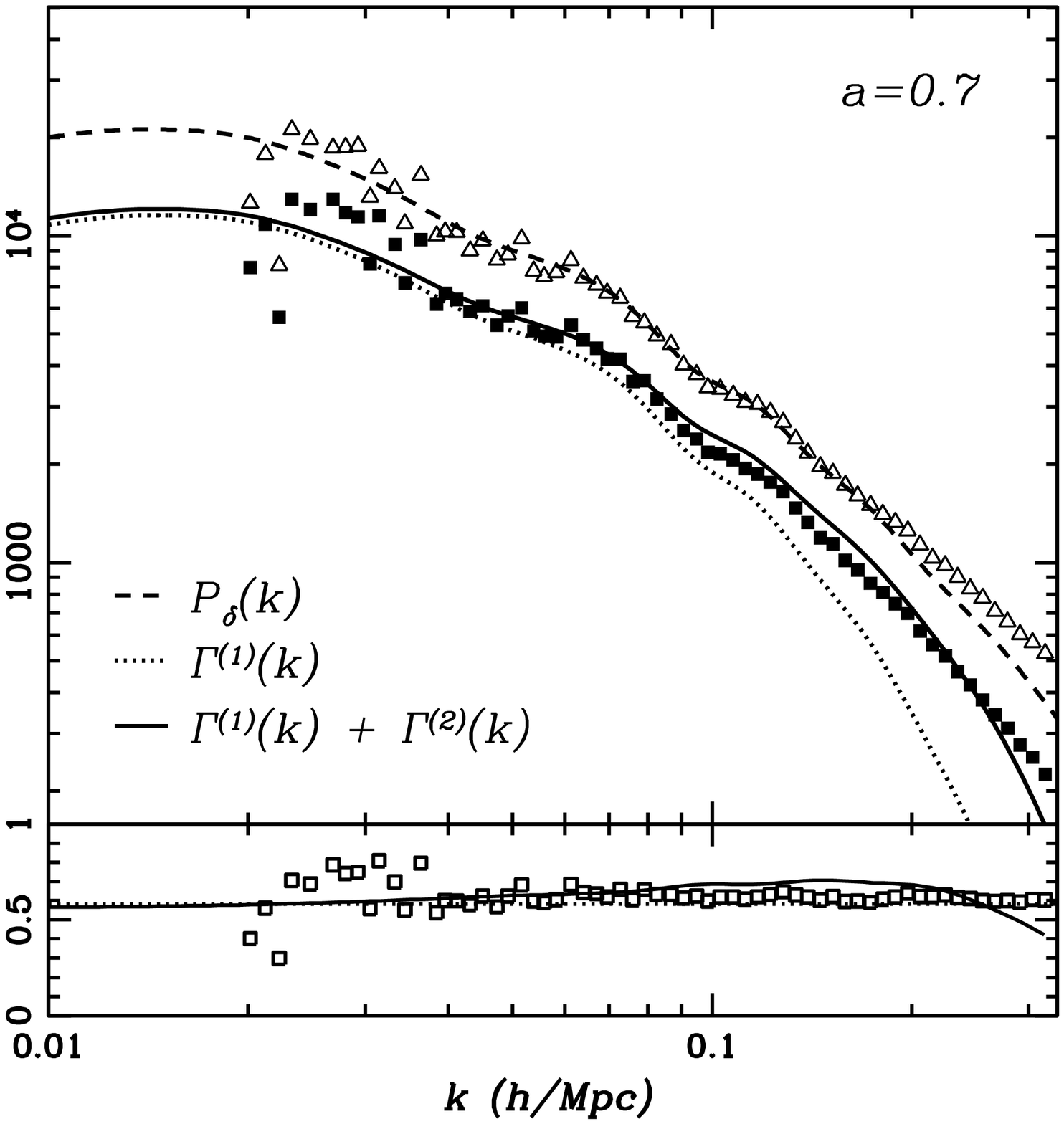}
\caption{\label{fig:pk_1} The nonlinear power spectrum of $A=\ln (1+\delta)$
at $a=1$ and $a=0.7$, with top-hat smoothing radius $R=3.7$\hmpc.
{\it Upper panels}:
Triangles and squares are the $\delta$ and $A$ power spectra measured 
from a simulation.
Solid lines include both the $ \Gamma^{(1)}_A$ and $\Gamma^{(2)}_A$ terms, 
and dotted lines include only the $ \Gamma^{(1)}_A$ term. 
Dashed lines are the linear power spectrum, damped at small scales with the NGP window function, to show the degree of nonlinearity of the $\delta$ power spectrum.  Note that we did not compute a perturbative prediction for the $\delta$ power spectrum in this paper.
{\it Lower panels}: the nonlinear power spectrum of $A$ divided by the linear 
density power spectrum.  Dotted horizontal lines indicate the value of 
the large scale bias.
In these figures, we use a different Fourier-transform convention; compared to previous sections, here power spectra are multiplied by $(2\pi)^3$.
}
\end{center}
\end{figure*}

In Fig.(\ref{fig:pk_1}), we illustrate the non-linear power spectra of
both $P_A(k)$ and $P_{\delta}(k)$ at $a=0.7$ and $a=1$.  The squares
and triangles are the $\delta$ and $A$ power spectra measured from
Coyote Universe simulation 0, which has 1024$^3$ particles in a cubic
1300 Mpc ($\sim1$\hgpc) box.  Its resolution is high enough that the
power spectrum is accurate to sub-percent level down to 1\ihmpc.  We
measure both power spectra on $128^3$-cell grids, using
nearest-grid-point (NGP) density assignment.  Even for the $A$ field,
the shot noise is negligible at this cell size \citep{NSS11}.  We do not correct
either measured power spectrum for the NGP pixel window function.  In
fact, such a correction would be ill-defined for the $A$ field.  The
dashed line shows the linear power spectrum, after applying the NGP window
function.  Note that we did not compute a perturbative prediction of
$P_{\delta}$ in this paper that would allow a fair shape
comparison to our perturbative prediction of $P_A$.

\begin{figure*}[!htp]
\begin{center}
\includegraphics[width=0.5\textwidth]{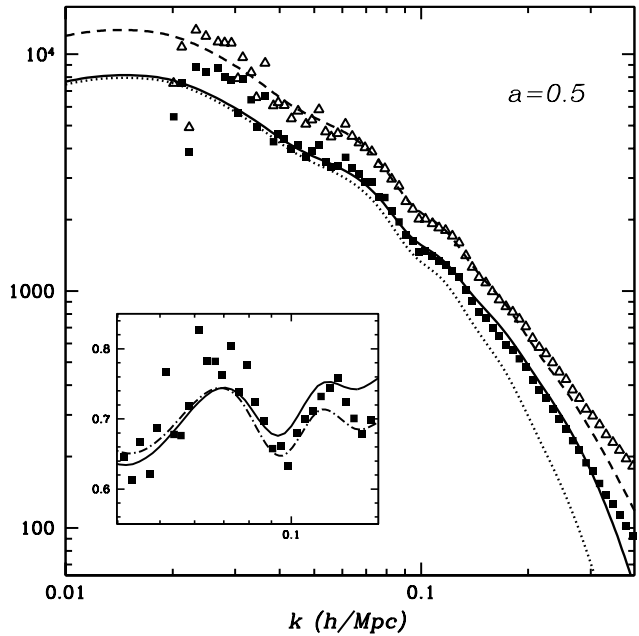}
\caption{\label{fig:pk_2} The nonlinear power spectrum of $A=\ln
  (1+\delta)$, at $a=0.5$, with top-hat smoothing radius $R=3.7$\hmpc.
  Linestyles and symbols the same as in Fig.\ \ref{fig:pk_1}.
  {\it Inset}: The solid line shows the BAO wiggles of $P_A(k)$, dividing out the
  \cite{EH98} `no-wiggle' power spectrum. The dot-dashed line presents
  the wiggliness of the linear $P_\delta(k)$, multiplied by a
  constant to line up with $P_A(k)$ at large scales.  }
\end{center}
\end{figure*}

In Fig.(\ref{fig:pk_2}), we show the same results at redshift $a=0.5$.
The calculation is quite accurate before it seems to break down at $k
\sim 0.31$\ihmpc.  In the inside panel, we show the BAO wiggles of both
$P_A(k)$ and linear $P_{\delta}(k)$, dividing out the \cite{EH98}
`no-wiggle' power spectrum.

The PT result of $P_A(k)$ (solid line) includes contributions from
$\Gamma^{(1)}_A$ as well as $\Gamma^{(2)}_A$.  Since the numerical
integration involving $\Gamma^{(2)}_A$ becomes extremely
time-consuming, we only calculate it up to one-loop order, while
$\Gamma^{(1)}_A$ is done to two-loop order.  The agreement with the
data measured from simulation is good, given the several
approximations we have made.  For $a=1$, the calculation breaks down
after $k\gtrsim 0.2$\ihmpc.  Also, the discrepancy with data between
$k=0.1$ and $k=0.2$\ihmpc\ peaks at over 10\%.  This is where
$\Gamma_A^{(2)}$ becomes dominant, but is only calculated up to
one-loop order.  The dotted line presents the $\Gamma_A^{(1)}$
contribution.  For higher redshifts, the differences between data and
theory become smaller.  At $a=0.7$, the prediction breaks down around
$k\sim 0.28$\ihmpc.  In the lower panel of each figure, we also plot
the ratio between $P_A(k)$ and the linear power spectrum, after
applying the pixel window function.

\section{Conclusion}
In this paper, we developed the cosmological perturbation theory 
for the log-density field $A=\ln(1+\delta)$.  In the context of
standard perturbation theory, we showed how different smoothing
scales can alter the amplitude of the $A$ power spectrum at large
scales. With the help of multi-point propagators developed in
renormalized perturbation theory, we constructed the building blocks
for the power spectrum of the $A$ field. In our formalism, each diagram
effectively includes infinitely many loop contributions.  The
Gaussian damping of the propagator for the $A$ field ensures that the
convergence of the series is well-controlled.  

We found that except for the large-scale bias, this $A$ propagator,
which quantifies the memory of initial conditions as a function of
scale, is similar to the $\delta$ propagator, damping at a similar
length scale.  This means that the memory of mode-by-mode initial
phases and amplitudes in the $A$ field is not much better than
in the $\delta$ field.  

Even with several approximations, our PT calculation for $P_A$
achieves good agreement with simulation measurements.  However,
further work is necessary to obtain results with sufficient precision
to resolve the question of whether the much-reduced nonlinearity in
the shape of $P_A$ compared to $P_\delta$ is understandable
perturbatively.

\acknowledgments 
We thank Katrin Heitmann and Adrian Pope for help
accessing the Coyote Universe simulations.  XW, MN and AS are grateful
for support from the W.M.\ Keck and the Gordon and Betty Moore
Foundations, and IS from NASA grants NNG06GE71G and NNX10AD53G, and
from the Pol\'anyi Program of the Hungarian National Office for
Research and Technology (NKTH).

\begin{appendix}
\newcommand{\appsection}[1]{\let\oldthesection\thesection
  \renewcommand{\thesection}{Appendix \oldthesection}
  \section{#1}\let\thesection\oldthesection}

\appsection{One-loop SPT}

\label{sec:1loop_spt}
In this appendix, we present the one-loop result of $P_A(k)$ in the context of
standard perturbation theory. Starting from Eq.\ (\ref{eqn:SPT_13}),
if one only considers the fastest-growing mode, $2P^{(13)}_A$ becomes
\begin{eqnarray}
\label{eqn:app_spt_1}
  2P^{(13)}_A (k, \eta) &=&  2P^{(13)}_{\delta}(k,\eta) W^2(kR) + 
  2 \sigma^2_R ~P_L(k,\eta) W^2(kR) -\nonumber \\
  && 4 \left [ \int d^3\vp ~ P_L(p,\eta) F^{(2)}(\vk,\vp) W(p R) W(|\vk+\vp| R) \right] P_L(k,\eta) W(k R),
\end{eqnarray}
where we have already expressed the four-point joint average in terms of the
power spectrum. $F^{(2)}(\vk_1,\vk_2)=\frac{5}{7}+\frac{1}{2}\mu 
\left(\frac{k_1}{k_2}+\frac{k_2}{k_1}\right)+\frac{2}{7} \mu^2$ is the second order 
perturbation kernel, where $\mu=\frac{\vk_1\cdot\vk_2}{k_1 k_2}$.
It can be further simplified by integrating out the angular part
of the third term of Eq.\ (\ref{eqn:app_spt_1})
\begin{eqnarray}
\label{eqn:app_spt_intg}
  \int d\Omega_p ~ \left[ 1+\frac{1}{2}\mu (\frac{p}{k}+\frac{k}{p}   )
  -\frac{2}{7}(1-\mu^2) \right] W(|\vk+\vp| R) ,
\end{eqnarray}
where $\mu$ is the cosine of the angle between the vectors $\vk$ and $\vp$.
To proceed, let us consider the spherical top-hat window function, 
which in Fourier space is defined as 
\begin{eqnarray}
 W_{th}(k R) = \frac{3}{(k R)^3} \left[ \sin(kR) - kR\cos(k R) \right].
\end{eqnarray}
For a given vector $\vk$, one can prove the following integration identities
\citep{B94a,B94,B96}:
\begin{eqnarray}
\label{eqn:app1}
 \int \frac{d\Omega_p}{4\pi} (1-\mu^2) W_{th}(|\vk+\vp|R) &=& \frac{2}{3} ~W_{th}{(kR)}
  W_{th}{(pR)}; \\
  \int \frac{d\Omega_p}{4\pi}\left [1+ \frac{1}{2}\mu \left(\frac{k}{p}+ \frac{p}{k}\right)\right]
  W_{th}(|\vk+\vp|R)& =& \frac{1}{2} \left[ \frac{\sin(pR)}{pR} W_{th}(kR) 
  + \frac{\sin(kR)}{kR} W_{th}(pR)  \right] =
   W_{th}(kR)  W_{th}(pR) \nonumber \\ && +\frac{1}{6} 
     \left[ W_{th}(kR)~ pR~W^{\pr}_{th}(pR)
   + W_{th}(pR)~ kR~W^{\pr}_{th}(kR)  \right] 
\label{eqn:app2}
\end{eqnarray}
Substituting Eqs.(\ref{eqn:app1} \& \ref{eqn:app2}) into Eq.\ (\ref{eqn:app_spt_intg}), 
one obtains
\begin{eqnarray}
  2P^{(13)}_A (k, \eta) &=&  2P^{(13)}_{\delta}(k,\eta) W^2(kR) - \frac{26}{21} 
  \sigma^2_R ~P_L(k,\eta) W^2(kR) - \frac{2}{3}  \biggl [\int dp~ 4\pi p^3 R~P_L(p,\eta) \nonumber \\ 
  && \times W_{th}(pR) W^{\pr}_{th}(pR) \biggr ]  
  P_L(k,\eta) W^2_{th}(k R) - \frac{2}{3} \sigma^2_R~ \left[kR W^{\pr}_{th}(kR)
  W_{th}(kR) \right] P_L(k,\eta), \nonumber \\
\end{eqnarray}
which reduces to Eq.\ (\ref{eqn:SPT_13simp}) if we define the derivative of $\sigma^2_R$
with respect to $R$, Eq (\ref{eqn:dsig}). 
As for $P^{(22)}_A(k)$, one can derive
\begin{eqnarray}
 P^{(22)}_A(k,\eta) &=&  P^{(22)}_{\delta}(k,\eta) W^2(kR) + 
  \int d^3\vp ~ P_L(p,\eta)P_L(|\vk-\vp|,\eta) 
    ~\biggl [\frac{1}{2} W^2(pR)  W^2(|\vk-\vp|R) \nonumber \\
    && - F^{(2)}(\vp, \vk-\vp) W(pR)W(|\vk-\vp|R)W(kR) \biggr ]
\end{eqnarray}

\appsection{Geometric Properties of the Top-hat Window Function}
\label{sec:tophat}
Integration identities Eq.\ (\ref{eqn:app1}) and Eq.\ (\ref{eqn:app2}) are nearly the
same as the one derived in the Appendix of \cite{B94a,B94}, except only one angular 
integration is involved. Following \cite{B94a,B94}, the top-hat window function can 
be decomposed as a sum of Bessel functions
\begin{eqnarray}
W_{th}(k) &=& 3\sqrt{\pi/2} k^{-3/2} J_{3/2}(k) \nonumber \\
W_{th}(|\vk+\vp|) &=& 3 \pi \sum_{m=0}^{\infty} \bigl (\frac{3}{2}+m \bigr)
(k p)^{-3/2} J_{3/2+m}(k) J_{3/2+m}(p)~ \frac{d}{d\mu} \mP_{m+1}(-\mu),
\end{eqnarray}
where $\mP_m$ is the Legendre polynomial of order $m$.  Therefore Eq.\ (\ref{eqn:app1})
can be expressed as
\begin{eqnarray}
 \int \frac{d\Omega_p}{4\pi} (1-\mu^2) W_{th}(|\vk+\vp|) = \frac{3\pi}{2}
\sum_{m=0}^{\infty} (\frac{3}{2}+m \bigr)
(k p)^{-3/2} J_{3/2+m}(k) J_{3/2+m}(p)
 \int_{-1}^1  d\mu  (1-\mu^2) ~ 
\frac{d}{d\mu} \mP_{m+1}(-\mu). \nonumber\\
\end{eqnarray}
Since  
$\int_{-1}^1  d\mu ~ (1-\mu^2)  ~\frac{d}{d\mu} \mP_{m+1}(-\mu)  = 
  \frac{4}{3} \delta_{0m}, $
Eq.\ (\ref{eqn:app1}) is proved.
\begin{eqnarray}
 \int \frac{d\Omega_p}{4\pi} ~(1-\mu^2) W_{th}(|\vk+\vp|) = 3\pi ~(kp)^{-3/2}
~ J_{3/2+m}(k) J_{3/2+m}(p) = \frac{2}{3}~ W_{th}(k) W_{th}(p).
\end{eqnarray}

For Eq.\ (\ref{eqn:app2}), by substituting $x=|\vk+\vp|R$, one gets
\begin{eqnarray}
\frac{3}{2} \int_{|\vk-\vp|R}^{|\vk+\vp|R} \frac{x~ dx}{kpR^2} ~\left[
1+ \frac{(k^2+p^2)(x^2/R^2-k^2-p^2 )}{4k^2p^2} \right]~\frac{\sin(x)-x\cos(x)}{x^3}
  = \frac{1}{2} \left[ W(kR)\frac{\sin(pR)}{pR} +
W(pR)\frac{\sin(kR)}{kR} \right]. \nonumber \\
\end{eqnarray}

\appsection{Two-loop Order of $\Gamma^{(1)}_A$}
\begin{figure}[!htp]
\begin{center}
\includegraphics[width=0.9\textwidth]{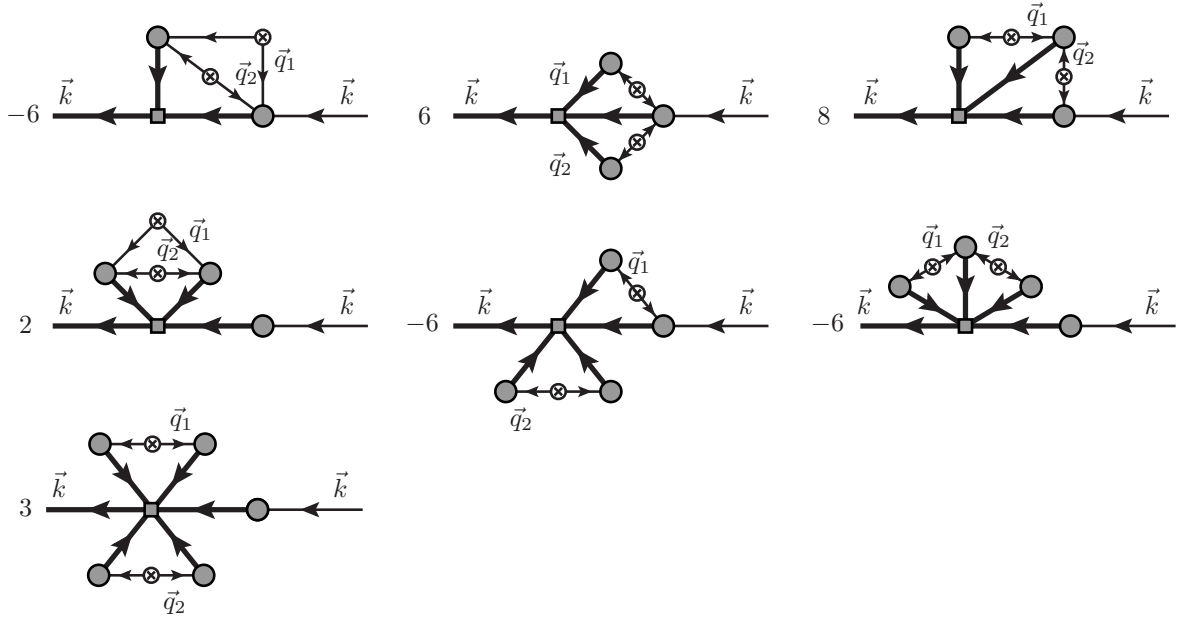}
\end{center}
\caption{\label{fig:GA1_2lp} Two-loop order of the two-point nonlinear propagator. 
}
\end{figure}

In this paper, we have calculated the two-point propagator $\Gamma_A^{(1)}$ up to 
two-loop order. Seven non-vanishing contributions are depicted in 
Fig.(\ref{fig:GA1_2lp}).
From these diagrams, one can write down all terms explicitly
\begin{eqnarray}
  \Gamma_{A~a}^{(\rm 1,~ 2-loop)}(\vk; \eta) 
  = \int d^3\vp_{12} ~ P_0(p_1)P_0(p_2) ~ 
  \left[ \sum_{i=1}^{7} ~  \mK^{(1,~\rm 2-loop)}_i \right ]
\end{eqnarray}
where 
\begin{eqnarray}
 \mK^{(1,~\rm 2-loop)}_1  &=& -6~ W(|\vp_1+\vp_2|R) W(|\vk+\vp_1+\vp_2|R)~
 \Gamma_{\delta}^{(2)}(-\vp_1, -\vp_2;\eta)  
 \Gamma_{\delta~a}^{(3)}(\vk, \vp_1,\vp_2;\eta),\nonumber \\ 
\mK^{(1,~\rm 2-loop)}_2  &=& 6~ W(p_1R) W(p_2 R)W(|\vk+\vp_1+\vp_2|R) ~ 
\Gamma_{\delta}^{(1)}(-\vp_1;\eta) \Gamma_{\delta}^{(1)}(-\vp_2;\eta)    
\Gamma^{(3)}_{\delta~a}(\vk, \vp_1, \vp_2;\eta),  \nonumber \\ 
 \mK^{(1,~\rm 2-loop)}_3  &=&8~ W(p_1R)W(|\vp_1+\vp_2|R) W(|\vk-\vp_2|R)~ 
 \Gamma_{\delta}^{(2)} (\vp_1, \vp_2;\eta) 
 \Gamma_{\delta}^{(1)}(-\vp_1;\eta) 
 \Gamma_{\delta~a}^{(2)} (\vk,-\vp_2;\eta) 
 \nonumber \\ 
 \mK^{(1,~\rm 2-loop)}_4  &=&2~W^2(|\vp_1+\vp_2|R)W(kR)~  
\Gamma_{\delta}^{(2)}(\vp_1,\vp_2;\eta) \Gamma_{\delta}^{(2)}(-\vp_1,-\vp_2;\eta)
 \Gamma_{\delta~a}^{(1)}(\vk;\eta) 
 \nonumber \\ 
 \mK^{(1,~\rm 2-loop)}_5  &=&-6~W(p_1R)W^2(p_2R)W(|\vk-\vp_1|R) 
 \Gamma_{\delta}^{(1)}(\vp_1;\eta) 
 \left[ \Gamma_{\delta}^{(1)}(\vp_2;\eta) \right]^2
 \Gamma_{\delta~a}^{(2)}(\vk,-\vp_1;\eta)  
 \nonumber \\ 
 \mK^{(1,~\rm 2-loop)}_6  &=&-6~W(kR)W(p_1R)W(p_2R)W(|\vp_1+\vp_2|R)~  
 \Gamma_{\delta}^{(2)}(\vp_1,\vp_2;\eta)  
 \Gamma_{\delta}^{(1)}(-\vp_1;\eta) 
 \Gamma_{\delta}^{(1)}(-\vp_2;\eta) 
 \Gamma_{\delta~a}^{(1)}(\vk;\eta) 
 \nonumber \\ 
 \mK^{(1,~\rm 2-loop)}_7  &=& 3~W^2(p_1R) W^2(p_2R) W(kR) ~
 \left[ \Gamma_{\delta}^{(1)}(\vp_1;\eta) \right]^2
 \left[ \Gamma_{\delta}^{(1)}(\vp_2;\eta) \right]^2
 \Gamma_{\delta~a}^{(1)}(\vk;\eta) 
\end{eqnarray}
\end{appendix}

\label{lastpage}


\begin{thebibliography}{}


\bibitem[Bernardeau(1994a)]{B94a}
  Bernardeau, F., 1994a, ApJ, 427, 51

\bibitem[Bernardeau(1994b)]{B94}
  Bernardeau, F., 1994b, ApJ, 433, 1

\bibitem[Bernardeau(1996)]{B96}
  Bernardeau, F., 1996, A\&A, 312, 11

\bibitem[Bernardeau et al.(2008)]{BCS08}
  Bernardeau, F., Crocce, M., Scoccimarro, R., 
  2008, PRD, 78, 103521

\bibitem[Coles \& Jones(1991)]{CJ91}
  Coles, P., Jones, B.,
  1991, MNRAS, 248, 1.

\bibitem[Crocce \& Scoccimarro(2006a)]{CS06a}
  Crocce, M., Scoccimarro, R.,
    2006, PRD, 73, 063519.

\bibitem[Crocce \& Scoccimarro(2006b)]{CS06b}
 Crocce, M., Scoccimarro, R.,
 2006, PRD, 73, 063520.

\bibitem[Crocce \& Scoccimarro(2007)]{CS07}
 Crocce, M., Scoccimarro, R.,
 2008, PRD, 77, 023533.

\bibitem[Eisenstein \& Hu(1998)]{EH98}
 Einsenstein, D.~J., Hu, W,
 1998, ApJ, 496, 605.

\bibitem[Hamilton(1985)]{H85} Hamilton, A.~J.~S.\ 1985, 
\apjl, 292, L35 

\bibitem[Heitmann et al.(2009)]{cu2} Heitmann, K., Higdon, 
D., White, M., Habib, S., Williams, B.~J., Lawrence, E., 
\& Wagner, C.\ 2009, \apj, 705, 156 

\bibitem[Heitmann et al.(2010)]{cu1} Heitmann, K., White, 
M., Wagner, C., Habib, S., \& Higdon, D.\ 2010, \apj, 715, 104 

\bibitem[Hubble(1934)]{H34} Hubble, E.\ 1934, \apj, 79, 8 

\bibitem[Kofman et al.(1994)]{K94} Kofman, L., 
Bertschinger, E., Gelb, J.~M., Nusser, A., 
\& Dekel, A.\ 1994, \apj, 420, 44 

\bibitem[Lawrence et al.(2010)]{cu3} Lawrence, E., 
Heitmann, K., White, M., Higdon, D., Wagner, C., Habib, S., 
\& Williams, B.\ 2010, \apj, 713, 1322 

\bibitem[{{Lewis} {et~al.}(2000){Lewis}, {Challinor}, \& {Lasenby}}]{camb}
{Lewis}, A., {Challinor}, A., \& {Lasenby}, A. 2000, \apj, 538, 473,
  \url{http://www.camb.info/}

\bibitem[Matsubara(2008)]{M08}
  Matsubara, T., 2008, MNRAS, 77, 063530

\bibitem[Neyrinck et al.(2009)]{NSS09}
  Neyrinck, M.~C., Szapudi, I., \& Szalay, A.~S.\
  2009, ApJ, 698, L90

\bibitem[Neyrinck et al.(2011)]{NSS11} Neyrinck, M.~C., 
Szapudi, I., \& Szalay, A.~S.\ 2011, \apj, 731, 116 

\bibitem[Pietroni(2008)]{P08}
  Pietroni, M., 
  2008, JCAP, 10, 36.

\bibitem[Smith et al.(2003)]{S03} Smith, R.~E., et al.\ 
2003, \mnras, 341, 1311

\bibitem[Springel et al.(2005)]{S05} Springel, V., et al.\ 
2005, \nat, 435, 629 

\bibitem[Szapudi \& Kaiser(2003)]{SK03}
  Szapudi, I., Kaiser, N., 
  2003, ApJ, 583, L1.

\bibitem[Taruya \& Hiramatsu(2008)]{TH08}
  Taruya, A., Hiramatsu, T., 
  2008, ApJ, 674, 617.







\end{thebibliography}
\end{document}